\documentclass[10pt,article,apl,aps,prx,superscriptaddress,amsmath,amssymb,floatfix,twocolumn,tightenlines]{revtex4-2}
\usepackage{amssymb}
\usepackage{natbib}
\usepackage{multirow}
\usepackage{bm}
\usepackage{amsmath}
\usepackage{graphicx}
\usepackage{epstopdf}
\usepackage{subfigure}
\usepackage{natbib}
\usepackage{epsfig}
\usepackage{amsfonts}
\usepackage{mathrsfs}
\usepackage{braket}
\usepackage{tabularx}
\usepackage{xcolor}
\usepackage[toc,page,title,titletoc,header]{appendix}
\usepackage{hyperref}
\hypersetup{colorlinks,allcolors=blue}
\usepackage{dsfont,amsthm,amsbsy}

\usepackage{fancyhdr}
\usepackage{lineno}

\begin{document}

\title{Imaging supermoiré relaxation in helical trilayer graphene}

\author{Jesse C. Hoke}
\thanks{These authors contributed equally}
\affiliation{Department of Physics, Stanford University, Stanford, CA 94305, USA}
\affiliation{Geballe Laboratory for Advanced Materials, Stanford, CA 94305, USA}
\affiliation{Stanford Institute for Materials and Energy Sciences, SLAC National Accelerator Laboratory, Menlo Park, CA 94025, USA}

\author{Yifan Li}
\thanks{These authors contributed equally}
\affiliation{Department of Physics, Stanford University, Stanford, CA 94305, USA}
\affiliation{Geballe Laboratory for Advanced Materials, Stanford, CA 94305, USA}
\affiliation{Stanford Institute for Materials and Energy Sciences, SLAC National Accelerator Laboratory, Menlo Park, CA 94025, USA}

\author{Yuwen Hu}
\thanks{These authors contributed equally}
\affiliation{Department of Physics, Stanford University, Stanford, CA 94305, USA}
\affiliation{Geballe Laboratory for Advanced Materials, Stanford, CA 94305, USA}
\affiliation{Stanford Institute for Materials and Energy Sciences, SLAC National Accelerator Laboratory, Menlo Park, CA 94025, USA}

\author{Julian May-Mann}
\affiliation{Department of Physics, Stanford University, Stanford, CA 94305, USA}

\author{Kenji Watanabe}
\affiliation{Research Center for Electronic and Optical Materials,
National Institute for Materials Science, 1-1 Namiki, Tsukuba 305-0044, Japan}

\author{Takashi Taniguchi}
\affiliation{Research Center for Materials Nanoarchitectonics,
National Institute for Materials Science, 1-1 Namiki, Tsukuba 305-0044, Japan}

\author{Trithep Devakul}
\affiliation{Department of Physics, Stanford University, Stanford, CA 94305, USA}

\author{Benjamin E. Feldman}
\email{bef@stanford.edu}
\affiliation{Department of Physics, Stanford University, Stanford, CA 94305, USA}
\affiliation{Geballe Laboratory for Advanced Materials, Stanford, CA 94305, USA}
\affiliation{Stanford Institute for Materials and Energy Sciences, SLAC National Accelerator Laboratory, Menlo Park, CA 94025, USA}

\begin{abstract}

In twisted van der Waals materials, local atomic relaxation can alter the underlying electronic structure. Characterizing lattice reconstruction and its susceptibility to strain is essential for understanding emergent electronic states, especially in multilayers where interference between moiré lattices yields larger supermoiré patterns whose energy is highly sensitive to local stacking. Here, we image spatial modulations in the electronic character of helical trilayer graphene which indicate relaxation into a superstructure of large domains with uniform moiré periodicity. We show that the supermoiré domain size is increased by strain and can be altered in the same device while preserving the local properties within each domain. Finally, we observe higher conductance at the domain boundaries, consistent with predictions that they host counter-propagating edge modes. Our work provides real-space visualization of moiré-periodic domains, reveals two independently tunable length scales, and demonstrates strain-engineering as a route toward designing correlated topological networks at the supermoiré scale.

\end{abstract}
\maketitle

\section*{Introduction} 

\begin{figure*}[t!]
    \renewcommand{\thefigure}{\arabic{figure}}
    \centering
    \includegraphics[width=1.99\columnwidth]{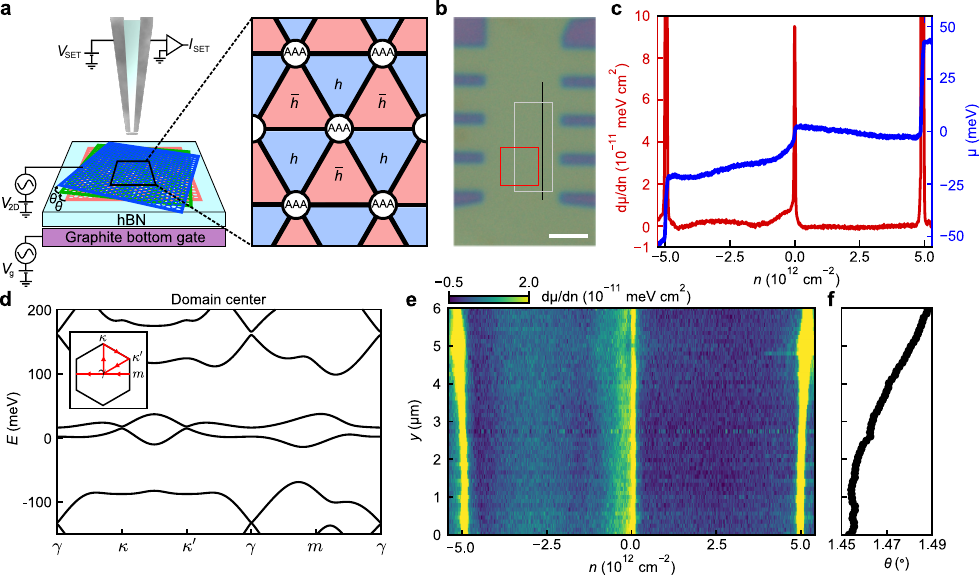}
    \caption{\textbf{Helical trilayer graphene (HTG) electronic structure}. \textbf{a}, Left: schematic of the measurement setup (see Methods). The HTG sample consists of three graphene layers consecutively twisted by $\theta \approx 1.5^\circ$ and is encapsulated by hexagonal boron nitride (hBN; top hBN not shown) with a graphite bottom gate. Right: schematic of the supermoiré domains predicted in HTG. The lattice relaxes into triangular domains (denoted as $h$ and $\bar{h}$) with uniform moiré periodicity, which are separated by domain walls (black lines) that intersect at AAA stacking sites (white circles). \textbf{b}, Optical micrograph of the HTG device. Scale bar (white): 2 $\mu$m. \textbf{c}, Inverse electronic compressibility d$\mu$/d$n$ (red) and chemical potential $\mu$ (blue) as a function of carrier density $n$ at temperature $T = 330$ mK. \textbf{d}, HTG band structure in a domain center for $\theta = 1.45^\circ$ and at displacement field $D=0$. Inset: the moiré Brillouin zone with the plotted path in momentum space indicated by red lines and arrows. \textbf{e}-\textbf{f}, Spatial line cut of d$\mu$/d$n$ as a function of $n$ at $T = 1.6$ K (\textbf{e}) and corresponding local twist angle (\textbf{f}) along the black line in \textbf{b}. 
} 
    \label{fig:htg}
\end{figure*}

The electronic properties of quantum materials depend sensitively on their lattice structure. Compared to conventional solids, the weak interlayer bonding in van der Waals systems provides additional freedom for atomic-scale relaxation driven by the competing energetics of different stacking configurations and elastic strain~\cite{nam2017lattice,carr2018relaxation,zhu2020modeling,nakatsuji2023multiscale}. As a result, local twists and strains can develop, enlarging areas with preferred stacking while shrinking disfavored arrangements~\cite{alden2013strain,huang2018topologically,yoo2019atomic,mcgilly2020visualization,kerelsky2021moireless,turkel2022orderly_disorder,craig2024local}. This has profound effects on both band structure and the resulting electronic states. For example, lattice relaxation in bilayers can generate new functionality, ranging from alternating ferroelectric domains driven by local symmetry breaking~\cite{yasuda2021stacking,vizner2021interfacial,woods2021charge,wang2022interfacial,weston2022interfacial} to solitons which host topologically protected boundary modes~\cite{martin2008topological,qiao2011electronic, ju2015topological} and can form networks of one-dimensional edge states~\cite{san2013helical,alden2013strain,huang2018topologically,yoo2019atomic,rickhaus2018transport,xu2019giant,mcgilly2020visualization,kerelsky2021moireless}. 
Additional tunability arises in twisted multilayers, as different interlayer shifts produce distinct electronic structure~\cite{khalaf2019magic,park2021tunable,turkel2022orderly_disorder,kim2022evidence,hao2021electric}, and interference between moiré patterns leads to a second, larger moiré-of-moiré (supermoiré) length scale which can also modify electronic behavior~\cite{xie2024strong,craig2024local,hesp2024cryogenic}.

Helical trilayer graphene (HTG), in which three graphene layers are consecutively twisted by equal angles $\theta$, provides an especially rich venue to explore the relationship between lattice relaxation, supermoiré structure, and electronic properties~\cite{mora2019flatbands,zhu2020twisted,mao2023supermoire,devakul2023magic,guerci2023chern,guerci2023nature,nakatsuji2023multiscale,yang2023multi,xia2023helical,popov2023magic,kwan2024strong,datta2024helical}. Nominally, HTG is quasi-periodic, as the moiré pattern formed by the first and second layers is incommensurate with that formed by the second and third layers. However, at small twist angles, HTG is predicted to relax into large triangular domains of moiré-periodic order~\cite{devakul2023magic,guerci2023chern,guerci2023nature,nakatsuji2023multiscale,yang2023multi}. This yields a uniform moiré wavelength $\lambda_{\rm{M}}\approx a/\theta$ within the individual domains, whose size is on the order of the supermoiré wavelength $\lambda_{\rm{SM}}\approx a/\theta^2$. Here $a$ is the lattice constant of graphene. Upon reconstruction, HTG has two in-equivalent types of domains (denoted as $h$ and $\bar{h}$ in Fig.~\ref{fig:htg}\textbf{a}) with different atomic stacking configurations. The domains locally break $C_{2z}$ symmetry, but map onto each other under a $C_{2z}$ transformation (Extended Data Fig.~1). Both domains locally host low energy bands that are separated from more dispersive bands by a large band gap.

Crucially, the Dirac cones of the three graphene layers, together with the superlattice gap induced by locally broken $C_{2z}$ symmetry, endows the low-energy bands of each valley with non-zero Chern numbers. Adjacent domains have opposite Chern number for a given valley such that HTG realizes a network of topological boundary modes when the domains are tuned to incompressible fillings~\cite{devakul2023magic,guerci2023chern}. Furthermore, near a magic angle $\theta\approx 1.8^{\circ}$, the topological bands become flat, making HTG a promising platform to investigate interaction-driven quantum anomalous Hall states~\cite{devakul2023magic,guerci2023chern,guerci2023nature,kwan2024strong,datta2024helical}. Indeed, a recent transport study of magic-angle HTG observed robust, but non-quantized anomalous Hall signals at both integer and fractional filling factors~\cite{xia2023helical}. It remains unknown whether the absence of quantization is related to the presence of domain walls. The above theoretical expectations and experimental signatures strongly motivate direct characterization of the lattice relaxation, its effect on electronic structure, and the role of domain walls, none of which have been experimentally addressed to date.

Here, we conduct single-electron transistor (SET) microscopy of HTG, which reveals periodic modulations of the local electronic structure at the supermoiré scale. Our measurements indicate that HTG relaxes into large domains of moiré-periodic order that locally break $C_{2z}$ symmetry. We demonstrate decoupling between moiré and supermoiré length scales, with strain capable of enlarging and reshaping domains without changing the local electronic character within them. We also observe increased conductivity along the network of domain walls when the domains are gapped, consistent with the prediction of helical edge modes. Our work clarifies the interplay between lattice reconstruction and strain, provides key insight for interpreting transport measurements of mesoscopic devices, and establishes that electronic states can be tuned on multiple independent length scales in moiré multilayers.

\section*{Electronic structure and lattice relaxation in HTG}

A schematic of the scanning SET measurement setup is illustrated in Fig.~\ref{fig:htg}\textbf{a} (see Methods), and an optical micrograph of the HTG device is shown in Fig.~\ref{fig:htg}\textbf{b}. We plot the inverse electronic compressibility d$\mu$/d$n$ as a function of carrier density $n$ (Methods) at a representative location within the sample in Fig.~\ref{fig:htg}\textbf{c} (red curve). A pair of prominent incompressible peaks occur near $n = \pm 5\times10^{12}$~cm$^{-2}$, which we attribute to the superlattice gaps at moiré filling factors $\nu=\pm 4$.  These gaps separate the flat bands near the charge neutrality point from the remote dispersive bands (Fig.~\ref{fig:htg}\textbf{d}). From the carrier densities of the superlattice peaks, we infer a local twist angle $\theta=1.45^{\circ}$ (Methods), and as we discuss below, the observation of only one pair of superlattice peaks indicates a single uniform moiré periodicity, despite the presence of two interlayer twist angles. The change in chemical potential across the flat bands is extremely small, $\Delta\mu_{\rm{FB}}\approx 20$~meV (Fig.~\ref{fig:htg}\textbf{c}, blue curve). The above characteristics match well to theoretical calculations of the HTG band structure~\cite{devakul2023magic,guerci2023chern,guerci2023nature,yang2023multi,kwan2024strong}.

We also observe a sharp incompressible peak at charge neutrality accompanied by pronounced electron-hole asymmetry. This is at odds with prior theoretical models, which predict gapless Dirac nodes at the charge neutrality point and nearly electron-hole symmetric bands~\cite{devakul2023magic,guerci2023chern}. The discrepancy can be reconciled by including momentum-dependent tunneling terms in the band structure calculation, which open a small band gap and qualitatively reproduce the asymmetric behavior (Supplementary Sec.~1), underscoring their importance for accurate modeling of the electronic structure. We also comment that while $\Delta\mu_{\rm{FB}}$ is smaller than that in magic-angle twisted bilayer~\cite{zondiner2020cascade,yu2023spin} and trilayer graphene~\cite{park2021tunable,liu2022isospin}, we do not observe correlation-driven gaps at intermediate filling factors. This suggests that the underlying single-particle bandwidth is a better proxy for strong interaction effects than $\Delta\mu_{\rm{FB}}$, which can be renormalized (broadened) by interactions~\cite{wong2020cascade,choi2021interaction}.

A spatial line cut along the black trajectory in Fig.~\ref{fig:htg}\textbf{b} provides the first indication of lattice relaxation in HTG. The electronic character is uniform (Fig.~\ref{fig:htg}\textbf{e}), and the superlattice peaks disperse only weakly, indicative of small local twist angle variations (Fig.~\ref{fig:htg}\textbf{f}). Importantly, we observe only a single pair of superlattice peaks throughout the spatial line cut. The same observation holds for the entire $3 \times 9\ \rm{\mu m}^2$ area of the device (Extended Data Fig.~2). This is notable because HTG has two interlayer angles, moiré systems are notoriously plagued by uncontrolled twist angle disorder \cite{uri2020mapping,lau2022reproducibility}, and spatial variations in each interlayer angle need not be correlated. One would then generically expect two distinct interlayer twist angles in a given location and thus two pairs of superlattice peaks, as observed in other twisted trilayer graphene systems~\cite{uri2023superconductivity,ren2024tunable}. 
However, based on the width in carrier density of the superlattice peaks, the interlayer angles would have to match to within $0.02^{\circ}$ throughout the device to produce a single pair of peaks, which we view as extremely unlikely. We also rule out that a large angle mismatch leads to a second superlattice peak outside the accessible range of carrier densities (Supplementary Sec.~2). Instead, we interpret the single pair of peaks as evidence for local relaxation that produces moiré-periodic order within the domains. 
While we cannot precisely determine the initial two twist angles prior to relaxation, theory predicts that relaxation is favored even when starting from two slightly mismatched angles, in which case the moiré periodicity is determined by an average of the interlayer angles~\cite{devakul2023magic,guerci2023chern,guerci2023nature,nakatsuji2023multiscale}.

\section*{Imaging supermoiré domains}

\begin{figure*}[t!]
    \renewcommand{\thefigure}{\arabic{figure}}
    \centering
    \includegraphics[width=1.99\columnwidth]{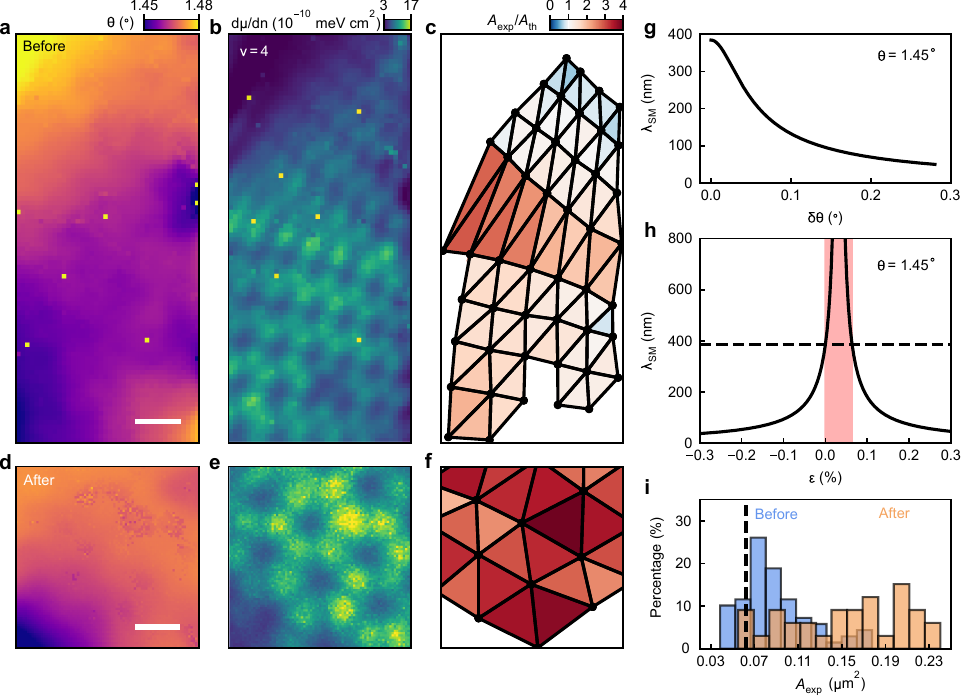}
    \caption{\textbf{Imaging supermoiré domains}. \textbf{a}, Spatial map of $\theta$ within the white box in Fig.~\ref{fig:htg}\textbf{b}. Scale bar: 500 nm. \textbf{b}, Value of d$\mu$/d$n$ at moiré filling factor $\nu = 4$ in the same area as in \textbf{a}. \textbf{c}, The local minima in \textbf{b} (indicated by black dots) correspond to the AAA stacking sites of HTG and form a triangular lattice with supermoiré wavelength $\lambda_\mathrm{SM}$. The color of each triangular supermoiré domain indicates the ratio of its experimentally observed area $A_\mathrm{exp}$ to the theoretically predicted area $A_\mathrm{th}$ in the absence of strain and for equal interlayer angles. \textbf{d}-\textbf{f}, Same as \textbf{a}-\textbf{c}, but in the red box in Fig.~\ref{fig:htg}\textbf{b} in a subsequent round of measurements after thermal cycling and other device changes. \textbf{g}, Dependence of $\lambda_\mathrm{SM}$ on the twist angle mismatch $\delta \theta$, where the two interlayer angles are $\theta \pm \delta \theta/2$ and $\theta = 1.45^\circ$. \textbf{h}, $\lambda_\mathrm{SM}$ as a function of global isotropic biaxial heterostrain $\epsilon$ on the middle layer for $\theta = 1.45^\circ$ and $\delta \theta = 0$. The pink shaded area denotes the narrow range for which $\lambda_\mathrm{SM}$ exceeds its predicted value in the absence of strain (black dashed lined). \textbf{i}, Histogram of the observed $A_\mathrm{exp}$ before and after thermal cycling. Black dashed line indicates $A_\mathrm{th}$ for $\theta = 1.45^\circ$ assuming $\epsilon=0$ and $\delta \theta = 0$. The ``After" histogram includes domains outside the field of view of panel \textbf{f} (Extended Data Fig.~6). All data measured at $T = 1.6$ K.
} 
    \label{fig:imaging}
\end{figure*}

\begin{figure*}[t!]
    \renewcommand{\thefigure}{\arabic{figure}}
    \centering
    \includegraphics[width=1.99\columnwidth]{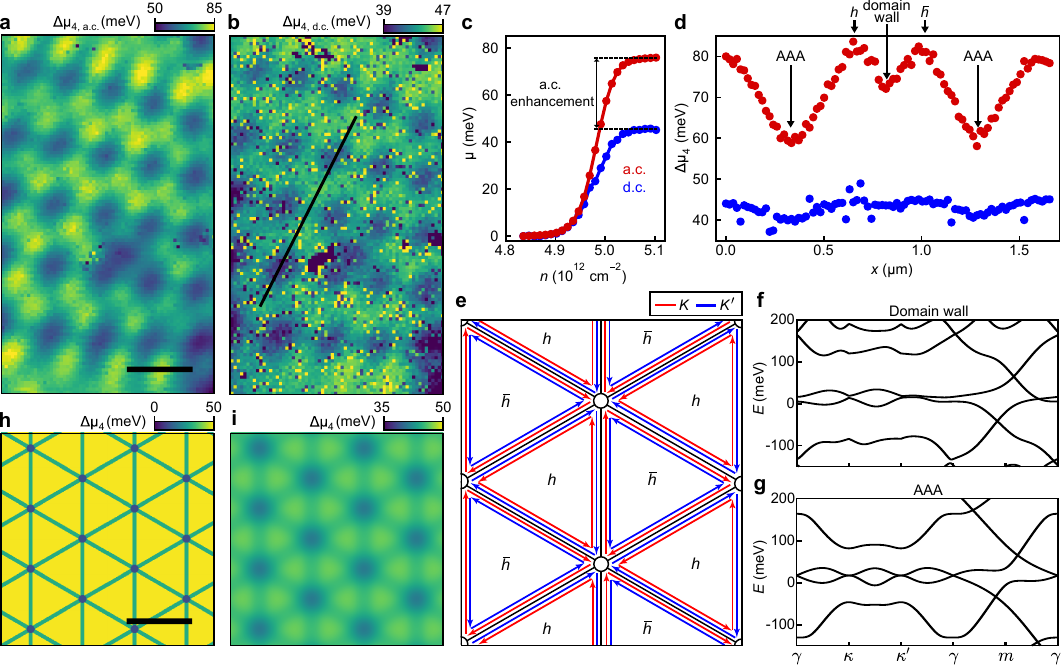}
    \caption{\textbf{Enhanced domain wall conductance}. \textbf{a-b}, High-resolution spatial maps of the change in chemical potential $\Delta \mu_{4}$ at $\nu = 4$ measured with a.c.~(\textbf{a}) and d.c.~(\textbf{b}) modalities. Scale bar: 500 nm. \textbf{c}, $\mu$ as a function of $n$ in the vicinity of $\nu = 4$, measured using a.c.~(red) and d.c~(blue) modalities. \textbf{d}, $\Delta \mu_{4}$ for a.c.~(red) and d.c.~(blue) modalities along the black trajectory from bottom left to top right in \textbf{b}. \textbf{e}, Schematic showing the predicted counter-propagating topological edge modes (for one spin) at $\nu=4$ along the boundaries of the $h$ and $\overline{h}$ domains in HTG. \textbf{f-g}, Band structure of HTG at a domain wall (\textbf{f}) and AAA site (\textbf{g}) at $D=0$. \textbf{h}, Theoretical spatial dependence of $\Delta {\mu}_{4}$ based on band structure calculations within the domains, at the domain walls, and at the AAA sites for $\theta=1.45^\circ$ and $D=0.45$ V/nm (see Supplementary Sec. 1c), with the domain size enlarged to match the experiment. Scale bar: 500 nm. \textbf{i}, Simulated spatial dependence of $\Delta {\mu}_{4}$ after accounting for the finite resolution of the SET tip. We assume a tip height of $h = 70$ nm and radius $R = 70$ nm (Supplementary Sec.~ 5). All data measured at $T = 1.6$ K.
} 
    \label{fig:conduct}
\end{figure*}

To further demonstrate the effects of lattice relaxation in HTG, we measure the spatial dependence of d$\mu$/d$n$ in the vicinity of the $\nu=4$ superlattice peak within a $\sim$10~$\rm{\mu m}^2$ area (white box in Fig.~\ref{fig:htg}\textbf{b}). The local twist angle (Fig.~\ref{fig:imaging}\textbf{a}) is remarkably homogeneous. In contrast, the peak value of d$\mu$/d$n$ at each location (Fig.~\ref{fig:imaging}\textbf{b}) exhibits approximately periodic spatial modulations that are not correlated with features in the twist angle map. The modulations occur on a length scale of a few hundreds of nanometers, significantly exceeding that of the moiré wavelength $\lambda_{\rm{M}} \approx 10$ nm. We therefore ascribe them to variations of the electronic bands in HTG at the supermoiré scale. Similar modulations with the same periodicity are also observed for all other measured incompressible states, both at zero magnetic field (Extended Data Fig.~3) and in a perpendicular magnetic field (Extended Data Fig.~4). 

The locations where d$\mu$/d$n$ at $\nu = 4$ has local minima (dark spots, Fig.~\ref{fig:imaging}\textbf{b}) form a triangular lattice. Surrounding these sites, the local maxima form a honeycomb structure (green, Fig.~\ref{fig:imaging}\textbf{b}), with narrow lines of lower d$\mu$/d$n$ separating adjacent maxima. This broadly agrees with the theoretical prediction that HTG relaxes into triangular $h$ and $\bar{h}$ domains, separated by domain walls where the local stacking order switches (Fig.~\ref{fig:htg}\textbf{a}, right)~\cite{devakul2023magic,guerci2023chern,guerci2023nature,nakatsuji2023multiscale,yang2023multi}. In particular, the triangular domains are predicted to be fully gapped at $\nu=4$, while their boundaries host gapless modes (see further discussion below). As a result, d$\mu$/d$n$ is maximized at the domain centers, while the gapless modes lead to enhanced compressibility at the domain walls, reaching a local minimum in d$\mu$/d$n$ where they meet at AAA stacking sites (Supplementary Sec.~1). We therefore associate the local minima in Fig.~\ref{fig:imaging}\textbf{b} with AAA stacking sites, and the local maxima with the centers of the $h$ and $\bar{h}$ domains.
We additionally note that the width in density $\Delta n$ of the superlattice peaks, which is affected by local variations in $\lambda_{\rm{M}}$, show similar modulations at the supermoiré scale. Namely, $\Delta n$ is maximized at the AAA sites and minimized within the domains (Extended Data Fig.~5). This is consistent with theoretical calculations~\cite{devakul2023magic} that show that while the domains of the relaxed HTG lattice structure are moiré-periodic, the AAA sites and domain walls are moiré-aperiodic.

Building on this qualitative understanding, we next assess the length scale of the supermoiré modulation more quantitatively. The experimentally observed distance between adjacent AAA sites is of the same order of magnitude as the supermoiré wavelength $\lambda_{\mathrm{SM}} \approx 380$ nm expected from the local twist angle of the device. However, variations in AAA site separation produce a range of supermoiré domain areas, defined as the triangular areas enclosed by adjacent AAA sites (Fig.~\ref{fig:imaging}\textbf{c}). We compare the experimentally observed areas $A_{\mathrm{exp}}$ to the theoretically predicted value $A_\mathrm{th}$ assuming equal interlayer angles and zero strain. Their ratio is encoded in the color of each triangle in Fig.~\ref{fig:imaging}\textbf{c}, which reveals substantial enhancement in the supermoiré domain area relative to the theoretical prediction over a large subset of the device.
We emphasize that this cannot be explained by twist angle mismatch $\delta\theta$ because $\lambda_{\mathrm{SM}}$ is maximal at $\delta\theta=0$ (Fig.~\ref{fig:imaging}\textbf{g}, Supplementary Sec.~2).

We instead attribute the enhanced supermoiré areas to global heterostrain. We hereafter refer to it only as strain, but note that it is distinct from local lattice relaxation. Such strain has been proposed to modulate domain size in HTG~\cite{devakul2023magic} and is ubiquitous in moiré systems~\cite{
yoo2019atomic,mcgilly2020visualization,kerelsky2021moireless,turkel2022orderly_disorder,lau2022reproducibility,hesp2024cryogenic}.
For simplicity, we first discuss isotropic biaxial heterostrain $\epsilon$ applied to the middle graphene layer, where $\epsilon>0$ denotes stretching. The supermoiré wavelength $\lambda_\mathrm{SM}$ is enhanced within a narrow range centered around a divergence at $\epsilon = 1 - \cos\theta$ (Fig.~\ref{fig:imaging}\textbf{h}). 
Heterostrain therefore provides a natural explanation for the increased domain sizes. We note that other strain configurations can also enlarge domains, and uniaxial or shear strain is required to account for the anisotropy of the observed supermoiré domains. We provide a more comprehensive discussion of other strain configurations in Supplementary Sec.~3.

Remarkably, we discover a dramatic reshaping of the supermoiré domains in subsequent measurements of the same device after thermal cycling and other changes (see Supplementary Sec.~4 for details). 
A second round of measurements in the area indicated by the red box in Fig.~\ref{fig:htg}\textbf{b} (largely overlapping with the bottom part of Fig.~\ref{fig:imaging}\textbf{a-b}) is shown in Fig.~\ref{fig:imaging}\textbf{d}-\textbf{e}. While the local twist angle is nearly identical, the supermoiré domain sizes substantially increase, up to $\sim$4 times the theoretically predicted area (Fig.~\ref{fig:imaging}\textbf{f}, \textbf{i}). Moreover, the supermoiré domains are more isotropic relative to those observed in the first round of measurements.

The areal enhancement of the domains, as well as the change in shape, can be attributed to redistribution of strain between measurements. In the new configuration, larger domains are favored to minimize regions associated with higher stacking energy, i.e.~domain walls and AAA sites~\cite{nam2017lattice,zhu2020modeling,nakatsuji2023multiscale,devakul2023magic}, with a relatively small cost of increased elastic energy from lattice relaxation and global heterostrain. We emphasize that this reshaping of supermoiré domains occurs with almost no change in the local twist angle (i.e.~moiré periodicity) or the band structure within the domains, as discussed in more detail at the end of the manuscript. Realistic values of heterostrain have little effect on the moiré wavelength, but can parametrically increase the supermoiré scale due to its divergence at moderate strain (Supplementary Sec. 3). Our work demonstrates a separation of length scales, providing new opportunities for strain engineering of supermoiré-scale networks without perturbing local moiré) physics.

\section*{Enhanced domain wall conductance}

We next address the electronic character of the domain walls in HTG, utilizing simultaneous, but independent measurements on a.c.~and d.c.~timescales with the SET (Methods). High-resolution spatial maps of the step in chemical potential $\Delta \mu_4$ at $\nu=4$ measured with each modality (Fig.~\ref{fig:conduct}\textbf{a}-\textbf{b}) show qualitatively similar supermoiré modulations to those in Fig.~\ref{fig:imaging}\textbf{b}: $\Delta \mu_4$ is maximized at the domain centers and minimized at the AAA sites. Quantitatively however, $\Delta \mu_{4, \mathrm{a.c.}}$ is larger than $\Delta \mu_{4, \mathrm{d.c.}}$, and it exhibits higher contrast. While both measurements should produce identical results when the sample conducts well, the a.c.~modality can exhibit spurious enhancement if sample resistance becomes large enough to make the $RC$ time constant comparable to the inverse a.c.~measurement frequency (Supplementary Sec.~5). Therefore, by studying local variations in the degree of a.c.~enhancement, we can qualitatively characterize the relative local conductivity of HTG.

The degree of a.c.~enhancement, defined in Fig.~\ref{fig:conduct}\textbf{c}, exhibits systematic spatial dependence across the supermoiré domain structure (Fig.~\ref{fig:conduct}\textbf{d}). It is largest in the center of the $h$ and $\bar{h}$ domains, is reduced at the domain walls, and is smallest at AAA sites. This indicates that the boundaries between domains are more conductive than their interiors (Supplementary Sec.~5). Theoretically, counter-propagating topological edge states with opposite Chern number for the two valleys are predicted along domain walls at $|\nu|=4$ in HTG (Fig.~\ref{fig:conduct}\textbf{e}). These edge modes manifest as remote Dirac cones within the superlattice gaps of local band structure calculations at the domain wall and AAA site (Fig.~\ref{fig:conduct}\textbf{f}-\textbf{g}, see Supplementary Sec. 1). While we only probe relative conductivity and thus do not directly address topology, our results are consistent with theoretical predictions.

To directly compare the data and theory, we calculate $\Delta\mu_4$ from the HTG band structure in domains, at domain walls, and at AAA sites, taking $\theta=1.45^\circ$ and accounting for the finite displacement field $D=0.45$ V/nm at $\nu=4$ (Supplementary Sec.~1). A spatial map of the theoretically calculated $\Delta\mu_4$ is shown in Fig.~\ref{fig:conduct}\textbf{h}. After accounting for the finite resolution of the SET, the simulated spatial dependence of $\Delta\mu_4$ produces a pattern that matches well with experiment (Fig.~\ref{fig:conduct}\textbf{i}, Supplementary Sec.~6).

\section*{Magnetic field dependence}

\begin{figure}[t!]
    \renewcommand{\thefigure}{\arabic{figure}}
    \centering
    \includegraphics[width=1\columnwidth]{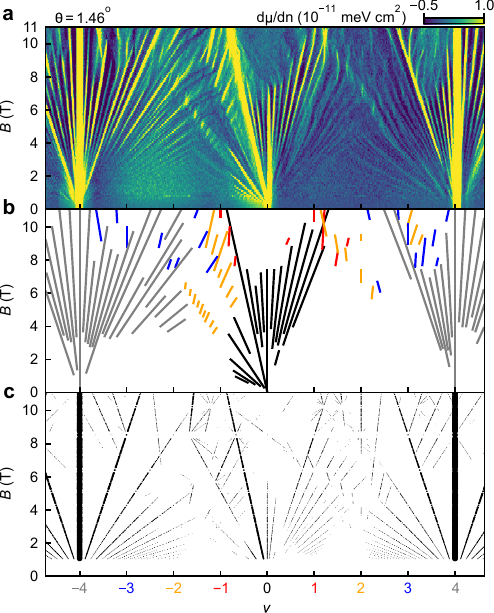}
    \caption{\textbf{Magnetic field dependence.} \textbf{a}, d$\mu$/d$n$ as a function of $\nu$ and perpendicular magnetic field $B$ in the center of a triangular domain at $T = 330$ mK. \textbf{b}, Wannier diagram of incompressible states identified from \textbf{a}. Black, red, orange, blue and grey states respectively correspond to states with integer zero-field intercepts at $|\nu| = 0, 1, 2, 3$, and $4$. \textbf{c}, Calculated Wannier-like diagram showing Hofstadter gaps for HTG at $\theta = 1.45^{\circ}$. The size of each dot reflects the gap magnitude.
} 
    \label{fig:landau}
\end{figure}

Finally, we discuss the behavior of HTG in a perpendicular magnetic field $B$. A Landau fan of d$\mu$/d$n$ as a function of $n$ and $B$ (Fig.~\ref{fig:landau}\textbf{a}) reveals an intricate pattern of quantum Hall and Hofstadter states. Each can be classified by an integer Chern number $C$ and zero-field intercept $\nu$. Within different regions of the Landau fan, we observe states that extrapolate back to every integer filling factor between $-4 \leq \nu \leq 4$, as shown in the Wannier diagram of Fig. \ref{fig:landau}\textbf{b}.

Qualitatively, the Landau fan is consistent with the low-energy band structure of HTG discussed above and can largely be captured by single-particle Hofstadter spectrum calculations (Fig.~\ref{fig:landau}\textbf{c}, Supplementary Sec.~1). We observe an electron-hole asymmetry due to momentum-dependent tunneling, and the most prominent states emanating from the charge neutrality point with $C=-4,-12, -20, \dots$ are consistent with the presence of two (weakly gapped) Dirac cones at the $\kappa$ and $\kappa'$ points (Fig.~\ref{fig:htg}\textbf{d}).

Multiple features in the Landau fan also indicate $C_{2z}$ symmetry breaking, which leads to distinct Hofstadter spectra for the $K$ and $K'$ valleys for $B \neq 0$. Specifically, the absence of large Hofstadter gaps at high fields near the charge neutrality point can be traced to crossing Hofstadter subbands that originate from different valleys (Extended Data Fig.~7, Supplementary Sec.~1). Similarly, the presence of Hofstadter states that extrapolate to intermediate integer $|\nu|=2$ are a consequence of the different Hofstadter spectra for each valley. We comment that these can arise without interaction effects. However, the weaker states with odd-integer Chern numbers or zero-field intercepts require spin splitting and likely reflect Hofstadter subband ferromagnetism~\cite{park2021flavour,choi2021correlation,saito2021hofstadter}. 

In Fig.~\ref{fig:landau}\textbf{a}, we observe neither evidence of moiré-of-moiré physics with states emerging from incommensurate filling factors~\cite{xie2024strong}, nor evidence for the presence of two distinct moiré periodicities~\cite{uri2023superconductivity, ren2024tunable}. This further supports the conclusion that HTG relaxes into large moiré-periodic domains. Moreover, the Landau fan in Fig.~\ref{fig:landau}\textbf{a} is generic across the entire sample (Extended Data Fig.~8) and the pattern of Hofstadter states was virtually identical in the second round of measurement (Extended Data Fig.~9), despite the significant rearrangement of the supermoiré domain shapes and sizes. These findings emphasize that although heterostrain can modulate the supermoiré pattern, the local electronic properties remain unchanged. 

\section*{Conclusion}
 
In conclusion, we directly imaged spatially modulated electronic states in HTG at supermoiré length scales, demonstrating a profound influence of lattice relaxation on electronic structure. The existence of moiré-periodic domains with conductive edge states at their boundaries has important implications for interpreting mesoscopic transport measurements. Our work strongly motivates future imaging and spectroscopic efforts to study the interactions between edge states and the degree to which they are protected from backscattering~\cite{huang2018topologically,randeria2019interacting}.
The observed separation of moiré and supermoiré length scales means that strain can adjust the supermoiré pattern without interrupting the local (moiré) electronic properties. This allows for design of reconfigurable spatially-modulated topological networks, including control over supermoiré symmetry, formation of fractional Chern mosaics~\cite{kwan2024fractional}, and potential for single-domain devices.
Due to the ideal quantum geometry of HTG at the magic angle, this may enable the realization of strongly-correlated topological states with quantized response~\cite{devakul2023magic,park2023observation,lu2024fractional}.
Lastly, similar effects are likely across a broad range of materials, such as twisted multilayers of transition metal dichalcogenides, 2D magnets, or superconductors, where intertwined topology, correlation, and magnetism can be engineered on the supermoiré scale.

\section*{Acknowledgements}
We thank Aaron Sharpe, Aviram Uri, Sergio de la Barrera, Li-Qiao Xia, Andrea Young, and Philip Kim for helpful discussions. This work was supported by the QSQM, an Energy Frontier Research Center funded by the U.S. Department of Energy (DOE), Office of Science, Basic Energy Sciences (BES), under Award \# DE-SC0021238. K.W. and T.T. acknowledge support from the JSPS KAKENHI (Grant Numbers 21H05233 and 23H02052) and World Premier International Research Center Initiative (WPI), MEXT, Japan. J.C.H. acknowledges support from the Stanford Q-FARM Quantum Science and Engineering Fellowship. Part of this work was performed at the Stanford Nano Shared Facilities (SNSF), supported by the National Science Foundation under award ECCS-2026822.

\section*{Author contributions}
J.C.H. and Y.L. fabricated the device. J.C.H., Y.L., and Y.H. conducted SET measurements. J.M.M., T.D., Y.L., Y.H., and J.C.H. performed theoretical calculations. T.D. and B.E.F supervised the project. K.W. and T.T provided hBN crystals. All authors contributed to analysis and writing of the manuscript. 

\section*{Competing interests}
The authors declare no competing interest. 

\section*{Data availability}
Data that support the findings in this study are available at https://doi.org/10.5281/zenodo.17365682.

\section*{Methods}
\subsection*{Device fabrication}

The HTG device was fabricated using standard dry transfer techniques with poly (bisphenol A carbonate)/polydimethylsiloxane (PC/PDMS) transfer slides. A monolayer graphene flake was cut into four pieces with a conductive atomic force microscope tip in contact mode. An exfoliated hBN flake (about 50 nm thick) was first used to pick up one of the monolayer graphene flakes, which served as a ``sacrificial layer" to prevent subsequent graphene layers from sliding on the hBN surface. We then used the hBN/monolayer graphene stack to sequentially pick the three remaining monolayer graphene flakes at the desired twist angles, with only a small overlap between them and the sacrificial layer to ensure a large bare HTG region remained. We next deposited the hBN/twisted graphene stack on top of a prefabricated bottom gate of few-layer graphite capped by hBN (17 nm), which was annealed in vacuum at 400~$^\circ$C for eight hours to ensure cleanliness of the surface. The full heterostructure was then patterned into a Hall bar geometry with standard electron beam lithography techniques followed by etching and metallization to form edge contacts. The sample at different stages of the fabrication process is shown in Extended Data Fig.~10.

\subsection*{Scanning SET measurements}
The SET tips were fabricated by evaporating aluminum onto the apex of a pulled quartz rod. The estimated diameter of each tip at its apex is approximately 100 nm. The tip is then brought to about 50-100 nm above the sample surface. The scanning SET measurements were performed in a Unisoku USM1300 scanning probe microscope system with a microscope head customized for scanning SET operation. All SET measurements are taken at $T = 330$ mK or $T = 1.6$ K and are explicitly noted in the figure captions. A bias voltage of $V_{\rm{SET}}= \ $2 mV is applied across the SET tip, and the current response $I_{\rm{SET}}$ is monitored. A voltage $V_{\rm{g}}$ is applied to the graphite bottom gate to tune the carrier density, while a voltage $V_{\rm{2D}}$ is directly applied to the HTG to minimize tip-induced doping.

The thermodynamic gap of an incompressible state is given by the corresponding step in the chemical potential $\Delta \mu$. $\Delta\mu$ is simultaneously and independently measured in two ways with the SET. We obtain d.c.~measurements of the chemical potential $\mu_{\rm{d.c.}}$ by tracking the voltage $V_{\rm2D}$ while $I_{\rm{SET}}$ is maintained at the maximum sensitivity point of the SET tip (i.e.,~the local electrostatic potential $\phi$ sensed by the tip is held constant). Following the relation $e\phi+\mu=eV_{\rm{2D}}$, the d.c.~chemical potential change can be read out as $\Delta\mu_{\rm{d.c.}}=e\Delta V_{\rm{2D}}$. Then the change of the chemical potential across the gap is,
\begin{equation}
\Delta \mu_\mathrm{d.c.} =  \mu_\mathrm{d.c.}(n_{+}) - \mu_\mathrm{d.c.}(n_{-}), 
\end{equation}
where $n_{+ (-)}$ are the densities immediately above and below the gap, respectively.

For the a.c.~measurements, a 2-5 mV peak-to-peak a.c.~excitation was applied to the sample ($\delta V_{\rm{2D}}$) and the graphite bottom gate ($\delta V_{\rm{g}}$), at respective frequencies of 823 Hz and 911.7 Hz, unless otherwise stated. The resulting modulations of the SET current $\delta I_{\rm{SET,2D}}$ and $\delta I_{\rm{SET,g}}$ are then measured by lock-in techniques at the corresponding frequencies. The inverse compressibility is obtained by ${\rm{d}}\mu/{\rm{d}}n=\frac{e^2}{C_{g}}\frac{\delta I_{\rm{SET,g}}/\delta V_{\rm{g}}}{\delta I_{\rm{SET,2D}}/\delta V_{\rm{2D}}}$, where $C_{g}$ is the geometric capacitance between the bottom gate and the sample. The a.c.~chemical potential change $\Delta\mu_{\rm{a.c.}}$ is obtained by integrating the inverse compressibility. The change of the chemical potential across the gap measured by a.c.~is,
\begin{equation}
\Delta \mu_\mathrm{a.c.} =  \mu_\mathrm{a.c.}(n_{+}) - \mu_\mathrm{a.c.}(n_{-}) = \int_{n_{-}}^{n_{+}} \left( \frac{{\rm{d}}\mu}{{\rm{d}}n} - \kappa^{-1}_B \right) \,\rm{d} \it{n}, 
\end{equation}
where $\kappa^{-1}_B$ is a small constant background that may need to be subtracted to set ${\rm{d}}\mu/{\rm{d}}n = 0$ where the d.c. chemical potential $\mu_{\rm{d.c.}}$ is flat. 

Mathematically, to extract $\Delta \mu$, we fit both $\Delta\mu_\mathrm{d.c.}$ and $\Delta\mu_\mathrm{a.c.}$ to a logistic function near $\nu = 4$. The fit takes the form:
\begin{equation}
\mu_\mathrm{fit}(n) = \mu_0 + \frac{\Delta \mu}{1+e^{-\gamma(n-n_0)}},
\end{equation}
where $\Delta \mu = \mu_\mathrm{fit}(n \rightarrow \infty) - \mu_\mathrm{fit}(n \rightarrow -\infty)$ is the fitted change in the chemical potential across an incompressible state, and $\mu_0$, $\gamma$, and $n_0$ are other fitting parameters whose specific values are irrelevant to extracting the gap size.

\subsection*{Twist angle determination}

The conversion from applied voltages to carrier density $n$ is determined by the geometric capacitance between the graphite bottom gate and the HTG sample: $n = C_{\rm{g}}(V_{\rm{g}}-V_{\rm{2D}})/e$, where $V_{\rm{g}}-V_{\rm{2D}}$ is the difference in the voltages applied to the bottom gate and the sample. The capacitance $C_{\rm{g}}$ between the bottom gate and sample is determined from the slopes of quantum Hall features emerging from charge neutrality in Landau fan measurements, which are quantized according to fundamental constants. This matches well to what is expected from geometrical considerations based on the hBN thickness measured by atomic force microscopy.

The average interlayer twist angle $\theta$ is determined based on the superlattice carrier density $n_{s} = 4/A \approx 8\theta^{2}/\sqrt{3}a^{2}$, which is found by a Gaussian fit to the superlattice peak in d$\mu$/d$n$. Here, $A$ is the moiré lattice unit cell area and $a = 0.246$ nm is the graphene lattice constant. We note that while we refer to a single angle $\theta$ locally throughout the manuscript, this should be interpreted as the angle corresponding to the single moiré periodicity after relaxation within the HTG domains, and it is not meant to imply that the two interlayer angles are identical before relaxation.

\bibliography{references.bib}

\setcounter{figure}{0}

\begin{figure*}[t!]
    \renewcommand{\thefigure}{ED\arabic{figure}}
    \centering
    \includegraphics[width=1.8\columnwidth]{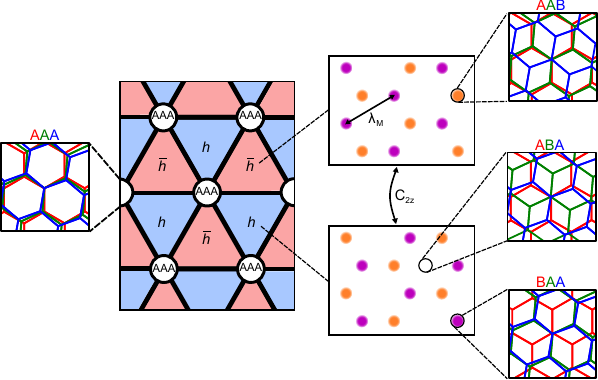}
    \caption{\textbf{Moiré-periodic order in HTG.} Schematic of the local stacking configurations within $h$ and $\bar{h}$ domains and AAA sites. The domains exhibit moiré-periodic order: a honeycomb lattice of alternating AAB and BAA stackings (orange and purple, respectively) with uniform moiré wavelength $\lambda_{\rm{M}}$, surrounded by local ABA stacking (white). The $h$ and $\bar{h}$ domains are related by a $C_{2z}$ transformation. The outermost schematics show the corresponding local real-space alignment of the bottom, middle, and top graphene layers in red, green, and blue, respectively. Here, A and B refer to the sublattices of the monolayer graphene honeycomb lattices.
} 
    \label{fig:moire}
\end{figure*}

\begin{figure*}[h!]
    \renewcommand{\thefigure}{ED\arabic{figure}}
    \centering
    \includegraphics[width=1.8\columnwidth]{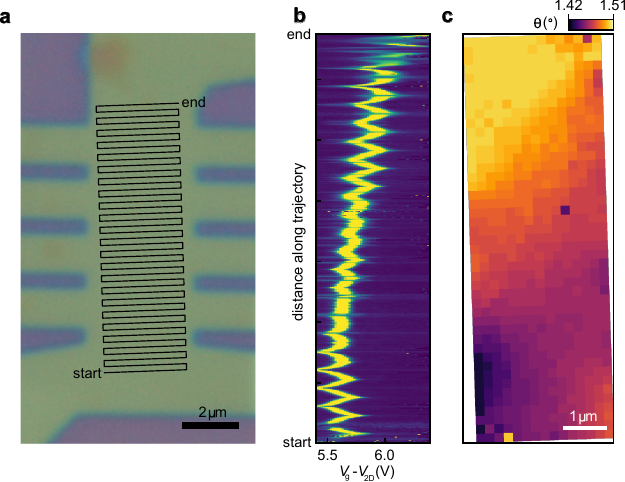}
    \caption{\textbf{Spatial dependence of the $\nu = 4$ superlattice peak.} \textbf{a}, Optical micrograph of the sample overlaid with the rasterized trajectory of the SET tip. 
    \textbf{b}, ${\rm{d}}\mu/{\rm{d}}n$ at $T= 1.6$ K in the vicinity of $\nu = 4$ along the trajectory in \textbf{a}. A single peak is observed throughout. \textbf{c}, Spatial map of $\theta$ determined from the location of the $\nu = 4$ peak in \textbf{b}. $\theta$ varies by less than $0.1^\circ$ across the entire area, highlighting the high degree of uniformity in the device.
} 
    \label{fig:trajectory}
\end{figure*}

\clearpage

\begin{figure*}[h!]
    \renewcommand{\thefigure}{ED\arabic{figure}}
    \centering
    \includegraphics[width=1.99\columnwidth]{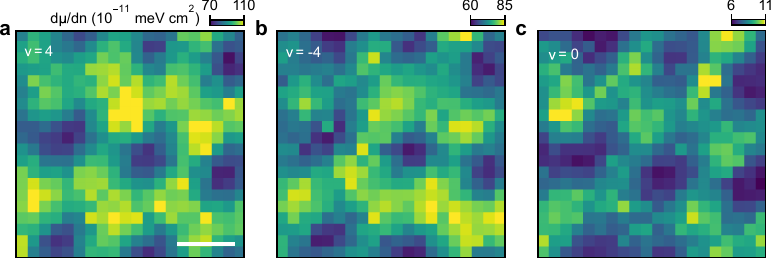}
    \caption{\textbf{Imaging supermoiré domains at different incompressible states at zero magnetic field.} \textbf{a}-\textbf{c}, Spatial dependence of ${\rm{d}}\mu/{\rm{d}}n$ in a $1 \times 1 \ \mathrm{\mu m}^2$ area of the sample at $\nu = 4$ (\textbf{a}), $\nu = -4$ (\textbf{b}) and $\nu = 0$ (\textbf{c}). Scale bar: 250 nm. Similar supermoiré modulations are visible for each incompressible state. This is consistent with the theoretical expectations that valley-contrasting domain boundary modes are present for each of these states. Data measured at $T = 1.6$~K for $|\nu|=4$ and at $T = 330$~mK for $\nu = 0$.
} 
    \label{fig:domains_0}
\end{figure*}

\begin{figure*}[h!]
    \renewcommand{\thefigure}{ED\arabic{figure}}
    \centering
    \includegraphics[width=1.8\columnwidth]{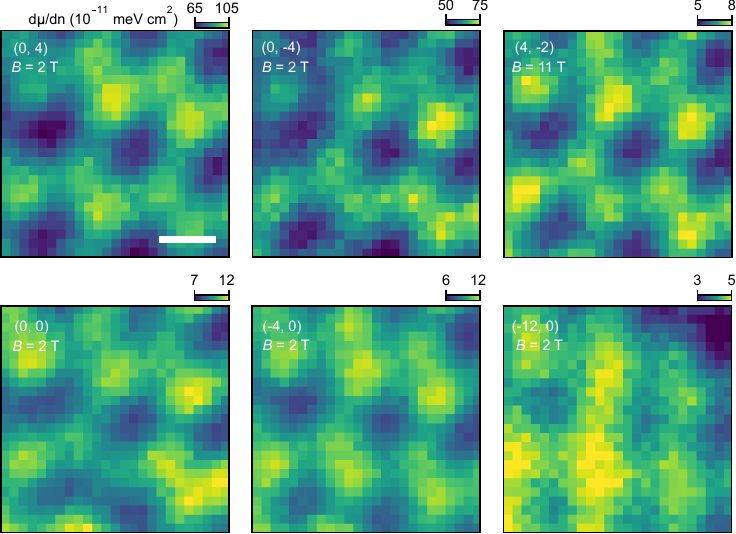}
    \caption{\textbf{Imaging supermoiré domains at different incompressible states at perpendicular magnetic field $B > 0$.} Spatial dependence of ${\rm{d}}\mu/{\rm{d}}n$ at $T=330$~mK in a $1 \times 1 \ \mathrm{\mu m}^2$ area of the sample for different incompressible states, respectively labeled by their Chern number $C$ and zero-field intercept $(C, \nu)$. Scale bar: 250 nm. Similar supermoiré modulations occur in all cases. This is again consistent with theoretical expectations that the valley-contrasting domain boundary modes are also stable in the presence of a magnetic field~\cite{datta2024helical}. 
} 
    \label{fig:domains_B}
\end{figure*}

\begin{figure*}[h!]
    \renewcommand{\thefigure}{ED\arabic{figure}}
    \centering
    \includegraphics[width=1.9\columnwidth]{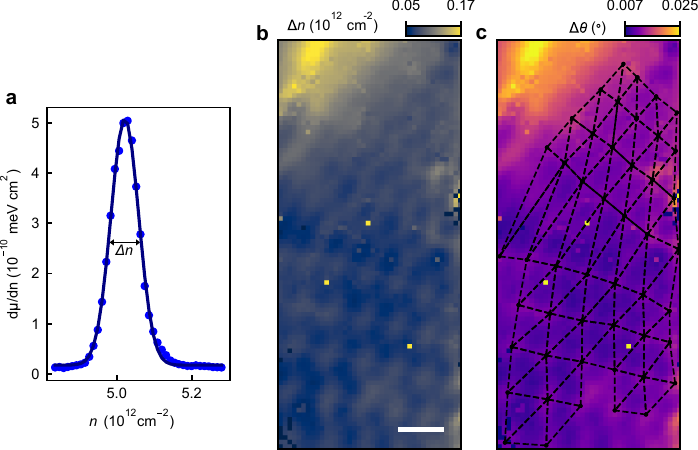}
    \caption{\textbf{Spatial dependence of the $\nu = 4$ superlattice peak width.} \textbf{a}, A representative plot of ${\rm{d}}\mu/{\rm{d}}n$ near $\nu = 4$. $\Delta n$ denotes the full width half maximum (FWHM) of the ${\rm{d}}\mu/{\rm{d}}n$ peak. Blue dots are data and the dark blue curve is a Gaussian fit. \textbf{b}, Map of the spatial distribution of $\Delta n$. Scale bar: 500 nm. \textbf{c}, The corresponding local twist angle disorder $\Delta \theta = \theta(n_{max}+\Delta n/2)-\theta(n_{max}-\Delta n/2)$, where $n_{max}$ is the density where the Gaussian peak is maximized and $\theta(n)$ is given by the equation in the ``Twist angle determination" subsection in the Methods. From the data, we estimate the median $\Delta \theta$ across the sample to be $\Delta \theta = 0.012^{\circ}$. This is (up to factors of order unity) an effective bound on the overall twist angle variability on length scales smaller than the $\sim 100$ nm spatial resolution of the SET probe. Notably, $\Delta \theta$ exhibits a periodic spatial dependence at the supermoiré scale. It is enhanced along domain boundaries and maximized at the AAA sites, while it is minimized within the domains. This observation is consistent with theoretical calculations of the relaxed HTG lattice structure: within the domains, the system is expected to be moiré-periodic and can be characterized by a single (relaxed) twist angle, whereas at the domain walls and AAA sites, the lattice structure is predicted to be moiré-aperiodic. The aperiodicity is expected to be larger at the AAA sites than at the domain walls~\cite{devakul2023magic}. The black dashed lines indicate the domain walls, while black dots at their intersections correspond to the AAA sites.
} 
    \label{fig:twist_disorder}
\end{figure*}

\begin{figure*}[h!]
    \renewcommand{\thefigure}{ED\arabic{figure}}
    \centering
    \includegraphics[width=1.6\columnwidth]{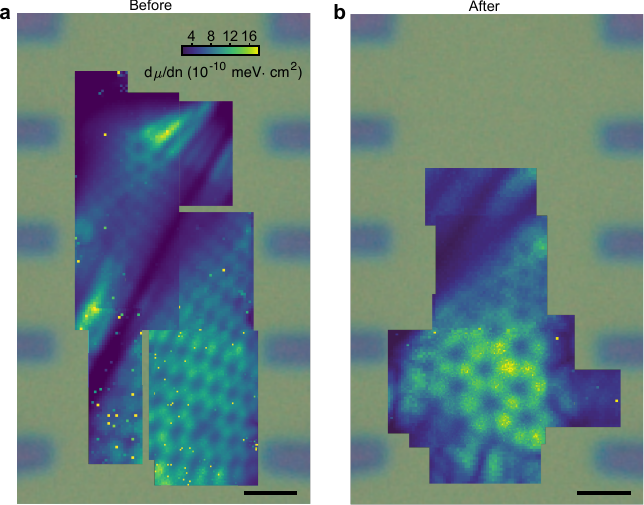}
    \caption{\textbf{Wider field of view of the supermoiré domains before and after thermal cycling and device changes.} \textbf{a}-\textbf{b}, Stitched images of d$\mu$/d$n$ at $\nu = 4$ during the first (\textbf{a}) and second (\textbf{b}) rounds of measurement. Scale bar: 1 $\mu$m. All data measured at $T=1.6$~K.
} 
    \label{fig:before_vs_after_imaging}
\end{figure*}

\begin{figure*}[h!]
    \renewcommand{\thefigure}{ED\arabic{figure}}
    \centering
    \includegraphics[width=1.8\columnwidth]{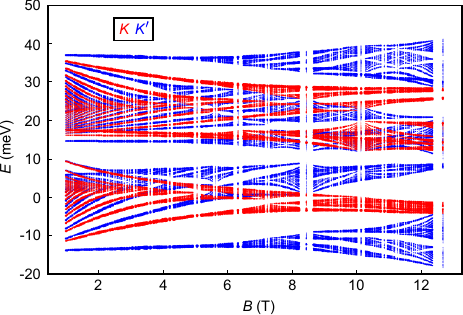}
    \caption{\textbf{Valley-resolved Hofstadter spectrum calculation.} Hofstadter spectrum of the $h$ domain. Colors indicate separate contributions from the $K$ (red) and $K'$ (blue) valleys, whose spectra are distinct due to $C_{2z}$ symmetry breaking. The Hofstadter spectrum of $\bar{h}$ domains is identical, except that $K$ and $K'$ are exchanged. 
}
    \label{fig:hoft_spectrum}
\end{figure*}

\begin{figure*}[h!]
    \renewcommand{\thefigure}{ED\arabic{figure}}
    \centering
    \includegraphics[width=1.9\columnwidth]{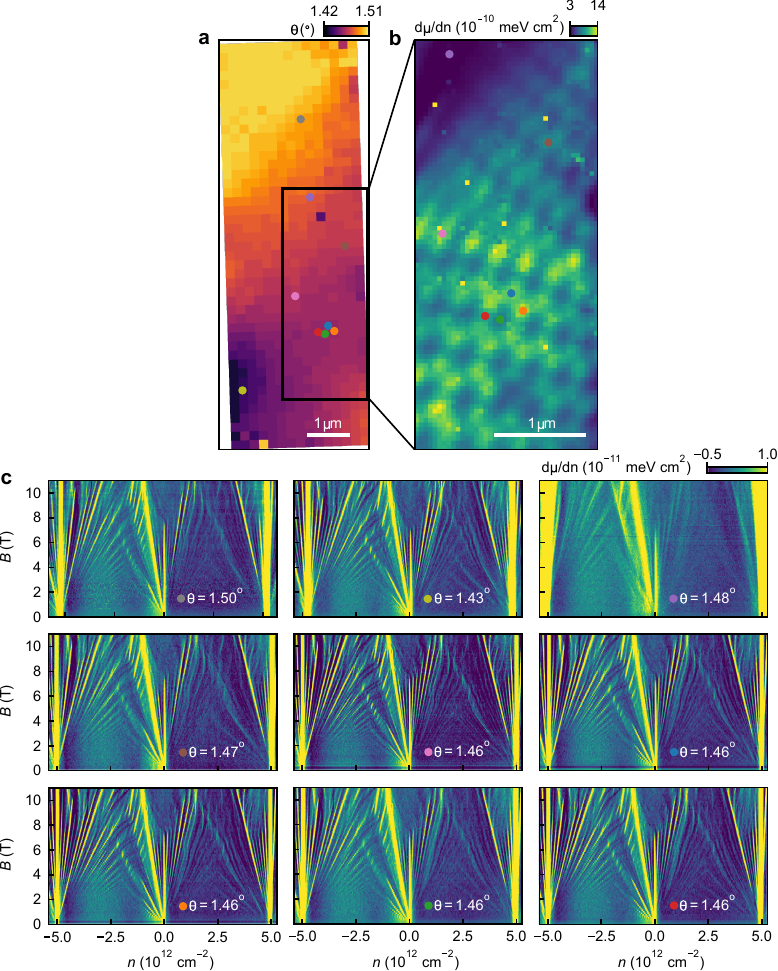}
    \caption{\textbf{Spatial dependence of Landau fan measurements.} \textbf{a}, Spatial map of $\theta$, reproduced from Fig.~\ref{fig:trajectory}. \textbf{b}, Peak ${\rm{d}}\mu/{\rm{d}}n$ at $\nu=4$ within the black box in \textbf{a}. The data are identical to that in Fig.~\ref{fig:imaging}\textbf{b}. \textbf{c}, ${\rm{d}}\mu/{\rm{d}}n$ at  $T = 330$ mK as a function of $n$ and $B$ in nine select locations (colored dots in \textbf{a} and \textbf{b}). Each Landau fan shows qualitatively similar behavior despite spanning different locations across several microns and representing a range of twist angles, differing apparent strain profiles, and different high symmetry locations within the supermoiré lattice.
} 
    \label{fig:landau_spatial}
\end{figure*}

\begin{figure*}[h!]
    \renewcommand{\thefigure}{ED\arabic{figure}}
    \centering
    \includegraphics[width=1.6\columnwidth]{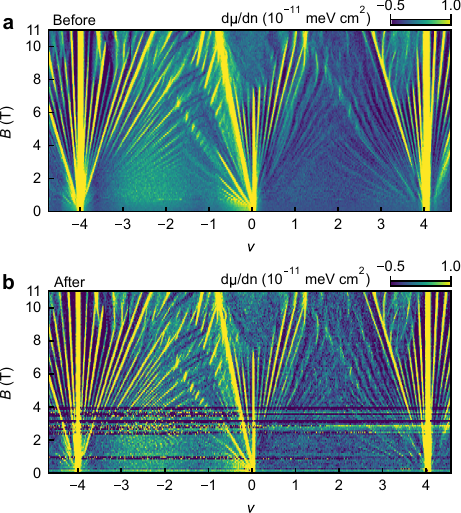}
    \caption{\textbf{Landau fan measurement before and after thermal cycling the device.} \textbf{a}-\textbf{b}, ${\rm{d}}\mu/{\rm{d}}n$ at $T = 330$ mK as function of $n$ and $B$, respectively measured in the center of a domain during the first (\textbf{a}) and second (\textbf{b}) round of measurements. The data in \textbf{a} are identical to Fig.~\ref{fig:landau}\textbf{a} and are reproduced here for ease of comparison.
} 
    \label{fig:after_fan}
\end{figure*}

\begin{figure*}[t!]
    \renewcommand{\thefigure}{ED\arabic{figure}}
    \centering
    \includegraphics[width=1.8\columnwidth]{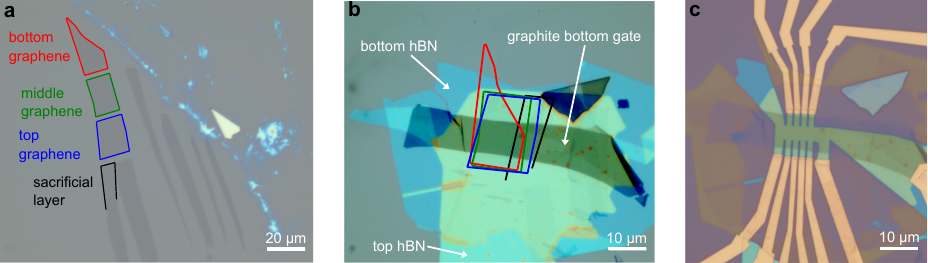}
    \caption{\textbf{Stacking and fabrication of the HTG device.} \textbf{a}, Optical image of the graphene flakes used in the HTG device after being cut into four pieces with an atomic force microscope. \textbf{b}, The final stack immediately after deposition onto a Si/SiO$_2$ substrate. A small overlap between the sacrificial layer (black) and the top graphene layer (blue) was used to prevent relative sliding between the hBN and top graphene layer. The relative angles between the bottom (red), middle (green), and top (blue) graphene flakes match closely to the target twist angles during stacking. \textbf{c}, Image of the final device after etching and metallization.
} 
    \label{fig:stacking}
\end{figure*}

\end{document}


\title{Supplementary Information for: \\ Imaging supermoiré relaxation in helical trilayer graphene}

\author{Jesse C. Hoke}
\thanks{These authors contributed equally}
\affiliation{Department of Physics, Stanford University, Stanford, CA 94305, USA}
\affiliation{Geballe Laboratory for Advanced Materials, Stanford, CA 94305, USA}
\affiliation{Stanford Institute for Materials and Energy Sciences, SLAC National Accelerator Laboratory, Menlo Park, CA 94025, USA}

\author{Yifan Li}
\thanks{These authors contributed equally}
\affiliation{Department of Physics, Stanford University, Stanford, CA 94305, USA}
\affiliation{Geballe Laboratory for Advanced Materials, Stanford, CA 94305, USA}
\affiliation{Stanford Institute for Materials and Energy Sciences, SLAC National Accelerator Laboratory, Menlo Park, CA 94025, USA}

\author{Yuwen Hu}
\thanks{These authors contributed equally}
\affiliation{Department of Physics, Stanford University, Stanford, CA 94305, USA}
\affiliation{Geballe Laboratory for Advanced Materials, Stanford, CA 94305, USA}
\affiliation{Stanford Institute for Materials and Energy Sciences, SLAC National Accelerator Laboratory, Menlo Park, CA 94025, USA}

\author{Julian May-Mann}
\affiliation{Department of Physics, Stanford University, Stanford, CA 94305, USA}

\author{Kenji Watanabe}
\affiliation{Research Center for Electronic and Optical Materials,
National Institute for Materials Science, 1-1 Namiki, Tsukuba 305-0044, Japan}

\author{Takashi Taniguchi}
\affiliation{Research Center for Materials Nanoarchitectonics,
National Institute for Materials Science, 1-1 Namiki, Tsukuba 305-0044, Japan}

\author{Trithep Devakul}
\affiliation{Department of Physics, Stanford University, Stanford, CA 94305, USA}

\author{Benjamin E. Feldman}
\email{bef@stanford.edu}
\affiliation{Department of Physics, Stanford University, Stanford, CA 94305, USA}
\affiliation{Geballe Laboratory for Advanced Materials, Stanford, CA 94305, USA}
\affiliation{Stanford Institute for Materials and Energy Sciences, SLAC National Accelerator Laboratory, Menlo Park, CA 94025, USA}

\maketitle

\section{1. Band structure and Hofstadter spectrum calculations}

\subsection{a. Theoretical model}

The theoretical single particle band structure of helical trilayer graphene (HTG) was calculated using a generalized Bistritzer-MacDonald continuum model~\cite{devakul2023magic,kwan2024strong,xia2023helical}. The starting point for this model is three graphene layers with lattice constant $a$. We explicitly define several quantities below to specify notation used in the following derivation. The atomic lattice vectors for layer $l$ are the columns of the matrices $\bm{A}_l = (1+\epsilon_l )\bm{R}(\theta_l) \bm{A}_0$. Here, 
\begin{equation}
    \bm{A}_0 = a \begin{pmatrix}
    1 & 1/2\\0 & \sqrt{3}/2
\end{pmatrix}, \phantom{==} \bm{R}(\theta_l) = \begin{pmatrix}
    \cos{\theta_l} & -\sin{\theta_l}\\ \sin{\theta_l} & \cos{\theta_l}
\end{pmatrix}
\end{equation}
are the lattice vectors of monolayer graphene and the two-dimensional rotation matrix, respectively. The angles $(\theta_1, \theta_2,\theta_3) = (\theta,0,-\theta)$ are the layer rotations for HTG, and $(\epsilon_1,\epsilon_2,\epsilon_3) = (0,1/\cos{\theta}-1,0)$ encodes 
a uniform dilation due to lattice relaxation~\cite{devakul2023magic}.
The reciprocal lattice vectors for layer $l$ are given by the columns of the matrix $\bm{G}_l = 2\pi \bm{A}^{-T}_l$. In terms of $\bm{G}_l$, the $K$ points of the three layers are $\vec{K}_l = \bm{G}_l (2/3,1/3)^T$. Due to the dilation, they lie on a straight line in momentum space, which gives rise to uniform moiré periodicity within the HTG domains. The Dirac points of the three layers can also be expressed in terms of the Dirac momentum $K_D = \frac{4\pi}{3 a}$, and the separation between Dirac nodes of adjacent layers $k_\theta = \frac{8\pi}{3 a} \sin(\theta/2)$. Explicitly, $(\vec{K}_1, \vec{K}_2, \vec{K}_3) = (K_D\cos(\theta)\hat{x} + k_\theta\hat{y}, K_D\cos(\theta)\hat{x}, K_D\cos(\theta)\hat{x}+ k_{-\theta}\hat{y})$.

Near these $K$ points, the continuum model is
\begin{equation}\begin{split}
H_K = \begin{bmatrix} v \vec{\sigma}_\theta\cdot [\vec{k} - k_{\theta}\hat{y}] + U & T_{1,2}(\vec{r}-\vec{d}_t,\vec{k}) & 0 \\ T^\dagger_{1,2}(\vec{r}-\vec{d}_t,\vec{k}) & v \vec{\sigma} \cdot \vec{k} & T_{2,3}(\vec{r}-\vec{d}_b,\vec{k}) \\ 0& T_{2,3}^\dagger(\vec{r}-\vec{d}_b,\vec{k}) & v \vec{\sigma}_{-\theta}\cdot [\vec{k} - k_{-\theta}\hat{y}]  - U \end{bmatrix}
\label{eq:BMHam}\end{split}\end{equation}
where $v$ is the Fermi velocity, $\vec{d}_{b(t)}$ is the relative interlayer shift of the moiré site of the bottom (top) graphene layer (values for high-symmetry supermoiré stackings are listed at the end of this section), $\vec{\sigma}_\theta = e^{-i\theta \sigma_z }(\sigma_x, \sigma_y)$, and $\sigma_{x,y,z}$ are Pauli matrices. The interlayer potential $U$ is related to the displacement field $D$ as $U = d_{\mathrm{int}} D/\epsilon_\perp$, where $d_{\mathrm{int}}$ is the interlayer spacing, and $\epsilon_\perp$ is the perpendicular dielectric constant.  The momentum $\vec{k}$ is measured relative to $\vec{K}_2$ (the $K$ point of the middle layer). 

The momentum-dependent tunneling matrix is
\begin{equation}\begin{split}
T_{l',l}(\vec{r},\vec{k}) = \sum^2_{j=0} e^{-i(\vec{Q}^j_{l',l
}) \cdot \vec{r}} w^j_{l',l}(\vec{k}) T_j    
\label{eq:tunnDef}\end{split}\end{equation}
where $\vec{r} \equiv -i \partial_{\vec{k}}$, thus $e^{-i \vec{Q}^j_{l',l}\cdot\vec{r}}$ is an operator that sends $\vec{k}\rightarrow\vec{k}-\vec{Q}^j_{l',l}$. Here, $T_j = \kappa \sigma_0 + \cos(\frac{2\pi}{3} j)\sigma_x + \cos(\frac{2\pi}{3} j)\sigma_y $, where $\kappa$ is the chiral ratio~\cite{tarnopolsky2019origin}. The tunneling wave vectors are $\vec{Q}^j_{l',l} = (\bm{G}_{l'} - \bm{G}_{l})\vec{m}_j$ where $\vec{m}_0 = (0,0)$, $\vec{m}_1 = (-1,0)$, $\vec{m}_2 = (-1,-1)$. Explicitly, the $\vec{Q}^j_{l',l}$ can be related to $k_{\theta}$: $\vec{Q}^j_{1,2} = \vec{Q}^j_{2,3}= \vec{q}_0-\vec{q}_j$ where $q_{j,x}+ i q_{j,y} = -i k_\theta e^{i \frac{2\pi}{3} j}$. Physically, the $j^{\text{th}}$ term in the sum describes a process where an electron in layer $l$ with momentum $\vec{k}+\vec{K}_2 + \bm{G}_{l}\vec{m}_j$ (relative to the $\Gamma$ point of the graphene layers) tunnels into a state on layer $l'$ with the same absolute momentum $\vec{k}-\vec{Q}^j_{l',l}+\vec{K}_2+\bm{G}_{l'}\vec{m}_j = \vec{k}+\vec{K}_2 + \bm{G}_{l}\vec{m}_j$.
The momentum dependence of $w^j_{l',l}$ reflects that, in real space, there are long range contributions to the interlayer tunneling. 

The explicit functional form of $w^j_{l',l}$ can be extracted from the microscopic interlayer tunneling amplitudes for the graphene $p_z$ orbitals. For two graphene sheets layers by a distance $d_{int}$, the interlayer tunneling amplitude is commonly parameterized as~\cite{koshino2015interlayer,slater1954simplified}
\begin{equation}
    t(\vec{r}) = - t e^{-(R-a_{cc})/r_0} \frac{r^2}{R^2} + t_\perp e^{-(R-d_{int})/r_0} \frac{d_{int}^2}{R^2},
\label{eq:realSpaceMicroTun}\end{equation}
where $a_{cc}$ is the nearest carbon-carbon distance in graphene, $r = |\vec{r}|$ is the in-plane distance between the two $p_z$ orbitals, and $R = \sqrt{r^2 + d_{int}^2}$ is the three-dimensional distance between the two $p_z$ orbitals.
Explicit values of the parameters will be given below. The tunneling matrix in Eq.~\ref{eq:tunnDef} is determined by the Fourier transform of $t(\vec{r})$, evaluated at the momentum (relative to $\Gamma$ of the graphene layers) where the tunneling occurs,
\begin{equation}
    w^j_{l-1,l}(\vec{k}) = |\bm{A}_0|^{-1} \int d^2 \vec{r} \exp\left ( - i (\vec{k}+\vec{K}_2 + \bm{G}_{l}\vec{m}_j)\cdot \vec{r}\right ) t(\vec{r})
\end{equation}
To proceed we make two observations. First, since $t(\vec{r})$ only depends on $|\vec{r}|$, $w^j_{l-1,l}(\vec{k})$ only depends on $|\vec{k}+\vec{K}_2 + \bm{G}_{l}\vec{m}_j|$. Second, for tunneling that occurs near the $K$ points of the graphene layers, $|\vec{k}+\vec{K}_2 + \bm{G}_{l}\vec{m}_j|$ is close to $K_D$. Based on this, we consider $w^j_{l-1,l}$ expanded to linear order in $|\vec{k}+\vec{K}_2 + \bm{G}_{l}\vec{m}_j| - K_D$,
\begin{equation}
    w^j_{l-1,l}(\vec{k}) \equiv w_0 + \frac{\rm{d}\omega}{\rm{d}\it{k}} (|\vec{k}+ \vec{K}_2 + \bm{G}_l \vec{m}_j| - K_D ).
    \label{eq:wk}
\end{equation}
The calculations in this work were performed using the following parameters: $\theta = 1.45^\circ$, $v = 10^6$~m/s, $a = 0.246$~nm, $\kappa = 0.7$, $d_{\mathrm{int}} = 0.33$~nm, $\epsilon_\perp = 2.73$, $\omega_0 = 110$~meV, and 
$\rm{d}\omega/\rm{d}\it{k} = \rm{-22.7}$~meV$\cdot$nm. 
The value of $\rm{d}\omega/\rm{d}\it{k}$ was determined using Eq.~\ref{eq:realSpaceMicroTun} with $t = 2.7$~eV, $t_\perp = 0.48$~eV, $a_{cc}= 0.142$~nm, and $r_0 = 0.0453$~nm. 

The AAA, $h$, $\bar{h}$ and domain wall regions correspond to interlayer shift $\vec{d}_t - \vec{d}_b = 0$, $\frac{1}{3}(\vec{a}_{\rm{M,2}} - \vec{a}_{\rm{M,1}})$, -$\frac{1}{3}(\vec{a}_{\rm{M,2}} - \vec{a}_{\rm{M,1}})$ and $\frac{1}{2}\vec{a}_{\rm{M,2}}$ respectively, where $\vec{a}_{\rm{M,1/2}} = \frac{4\pi}{3k_{\theta}}(\pm \frac{\sqrt{3}}{2},\frac{1}{2})$ are the moiré lattice vectors. The resulting band structures at displacement field $D=0$ inside a domain, on the domain wall, and on an AAA site are plotted in Fig.~1\textbf{d}, Fig.~3\textbf{f}, and Fig.~3\textbf{g}, respectively. This displacement field is appropriate near the charge neutrality point. Here and in all band structure plots, zero energy $E$ corresponds to the energy of the charge neutrality point of monolayer graphene.

\subsection{b. Effect of momentum-dependent tunneling}

\begin{figure}[t!]
    \renewcommand{\thefigure}{S\arabic{figure}}
    \centering
    \includegraphics[width=0.72\columnwidth]{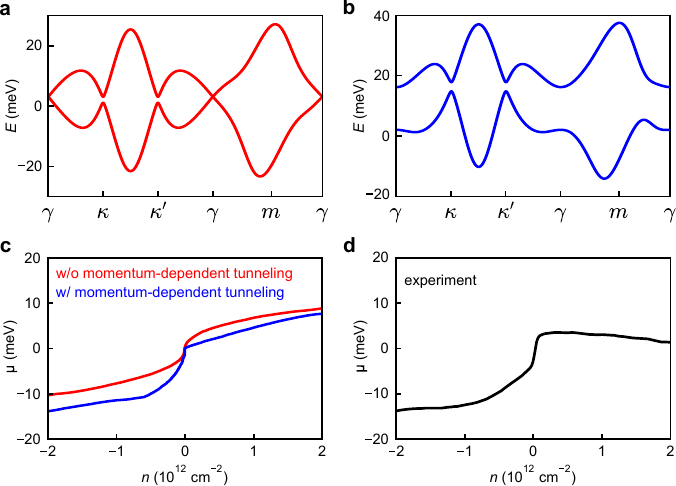}
    \caption{\textbf{Effect of momentum-dependent tunneling.} \textbf{a}-\textbf{b}, Helical trilayer graphene (HTG) band structure for $\theta = 1.45^\circ$ near the charge neutrality point (CNP) without momentum-dependent tunneling (\textbf{a}) and upon its inclusion (\textbf{b}) at the domain center and a perpendicular displacement field $D=0$. \textbf{c}, Theoretically determined chemical potential $\mu$ as a function of carrier density $n$ from the band structures without (red) and with (blue) the momentum-dependent tunneling term. The chemical potential is defined such that $\mu(n=0)=0$ for each curve. \textbf{d}, Corresponding experimental measurement of $\mu(n)$. 
} 
    \label{fig:D0_zoomin}
\end{figure}

To clarify the impact of the momentum-dependent tunneling terms, we compare the band structure with and without them, as well as the corresponding chemical potential $\mu$ as a function of carrier density $n$. A zoom-in on the low-energy band structure of a HTG domain without these terms (but with lattice relaxation) is shown in Fig.~\ref{fig:D0_zoomin}\textbf{a}. The resulting $\mu(n)$ is smooth and nearly electron-hole symmetric (red curve, Fig.~\ref{fig:D0_zoomin}\textbf{c}). In contrast, including momentum-dependent tunneling breaks the particle-hole symmetry and gaps out the Dirac node at $\gamma$ (Fig.~\ref{fig:D0_zoomin}\textbf{b}). The resulting $\mu(n)$ displays a sharp cusp at charge neutrality (blue curve, Fig.~\ref{fig:D0_zoomin}\textbf{c}) and shows excellent agreement with the experimental data (Fig.~\ref{fig:D0_zoomin}\textbf{d}). The difference in slope of $\mu(n)$ between theory and experiment may reflect additional interaction effects not included in the single-particle calculation.

\subsection{c. Effects of remote hopping and other intralayer terms}

In this subsection, we analyze the effects of various additional terms in the continuum model on the band structure in the domain center of HTG. Like the momentum-dependent tunneling term, these additional terms can also lead to particle-hole asymmetry and/or gaps at one or more Dirac nodes. Here we quantify their impact on the band structure.

Specifically, we consider the role of two types of terms that are neglected by the Bistritzer-MacDonald model: (1) a remote interlayer hopping term between the outer two layers, and (2) other higher-order intralayer terms. In order to isolate their contributions to particle-hole asymmetry, we turn off the momentum-dependent tunneling term ($\mathrm{d}\omega/\mathrm{d}k=0$ and Pauli twists ($\vec{\sigma}_{\theta}=\vec{\sigma}$), in which case the Hamiltonian in Eq.~\ref{eq:BMHam} has an exact inversion-particle-hole symmetry~\cite{devakul2023magic}. This spectrum has exact $E(\vec{k})\leftrightarrow -E(-\vec{k})$ symmetry and three protected Dirac cones at $\gamma$, $\kappa$, and $\kappa^\prime$ in the mini Brillouin zone. We refer to this Hamiltonian as the ``bare'' model, and consider the effects of additional terms on top of it.

\subsubsection{Remote interlayer hopping}
\begin{figure}[t!]
    \renewcommand{\thefigure}{S\arabic{figure}}
    \centering
    \includegraphics[width=0.4\columnwidth]{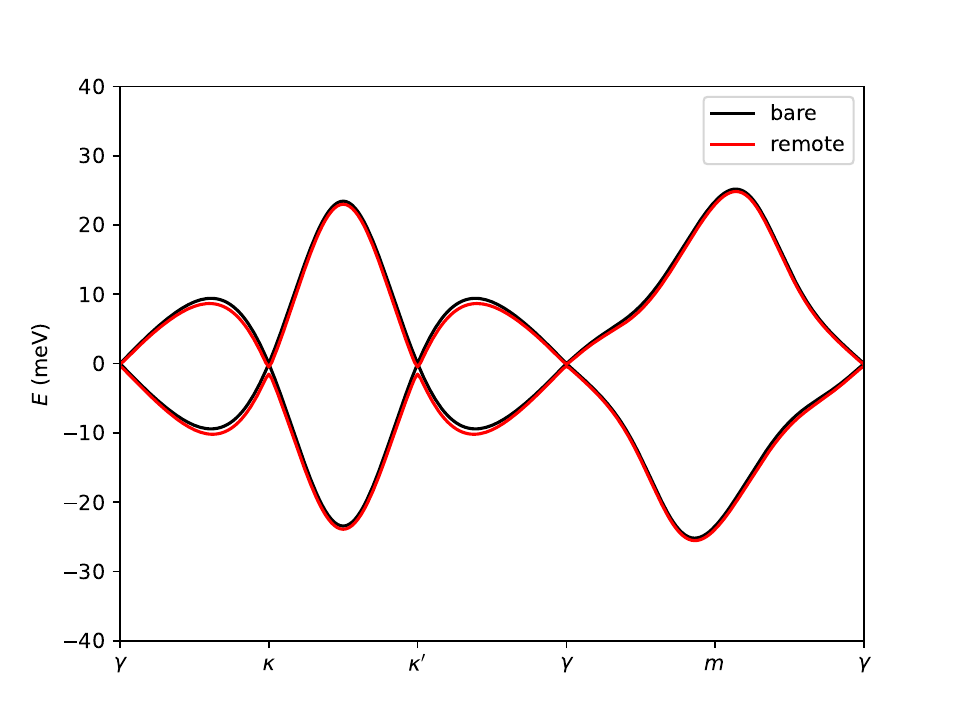}
    \caption{\textbf{Effect of the remote interlayer hopping.} Zoom-in of the band structure near the CNP with the remote interlayer hopping term (red) and with the bare model (black). The remote interlayer hopping term opens small gaps on the Dirac cone, weakly breaking the particle-hole symmetry in the band structure, but not by an amount that can explain the experimental results.
} 
    \label{fig:bare_remote}
\end{figure}

We first consider a remote interlayer hopping term between layers $1$ and $3$.
To estimate the magnitude of this term, we notice that the ratio of the next-adjacent layer hopping to the adjacent layer hopping is $\lambda\approx -0.05$~\cite{castro2009electronic,zhang2010band}. Setting the relative moiré shifts to be $\vec{d}_t=-\vec{d}_{b}$, taking this term into account simply amounts to adding a $\lambda T_{1,3}(\vec{r},\vec{k})$ term to the upper right corner of the Hamiltonian matrix in Eq.~\ref{eq:BMHam}, along with its hermitian conjugate. In Fig.~\ref{fig:bare_remote}, we show the bare model band structure alongside the bands with remote hopping taken into account. This term breaks the exact $E(\vec{k})\leftrightarrow -E(-\vec{k})$ symmetry and produces a small gap at the Dirac points, but the overall magnitude of this effect is very small (especially compared to the effect of the momentum-dependent tunneling term). We thus conclude that remote interlayer hopping terms alone cannot explain the experimental data.

\subsubsection{Higher-order intralayer terms}
We now consider the effect of higher-order intralayer terms. We follow closely the results in Ref.~\cite{kang2023pseudomagnetic}, with the set of parameters from~\cite{moon2012energy}.
There are two types of corrections: (i) terms arising at second-order in the momenta, and (ii) terms arising from an inhomogeneous strain field.

\begin{figure}[t!]
    \renewcommand{\thefigure}{S\arabic{figure}}
    \centering
    \includegraphics[width=0.4\columnwidth]{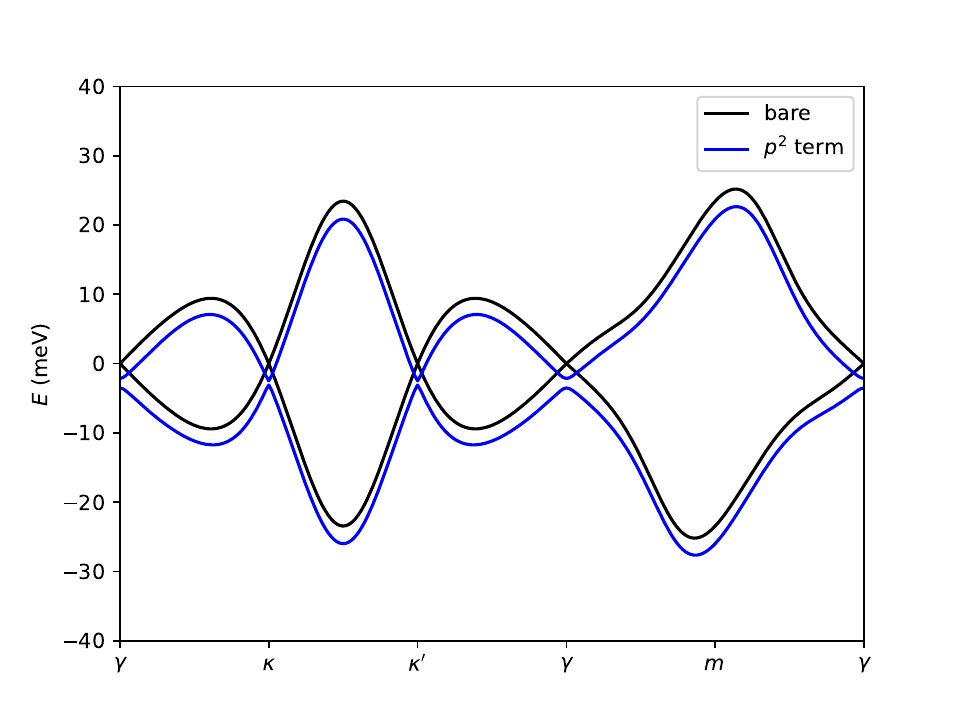}
    \caption{\textbf{Effect of the terms arising at second-order in the momenta.} Zoom-in of the band structure near the CNP with the terms arising at second-order in the momenta (blue) and with the bare model (black). This term opens small gaps on the Dirac cone and shifts the entire spectrum in energy, breaking the particle-hole symmetry in the band structure.
} 
    \label{fig:bare_p2}
\end{figure}

The effect of the first type in the Hamiltonian is taken into account by 
\begin{equation}
\delta H_{1}^{\ell}(\vec{k}) = \beta_0 |\vec{k}|^2 \sigma_0 + \beta_1 ((k_x^2-k_y^2)\sigma_x-2k_xk_y\sigma_y)
\end{equation}
with $\beta_0/a^2=-0.18$ eV and $\beta_1/a^2=-0.37$ eV.
The Pauli matrices $\sigma_x$ and $\sigma_y$ act on the sublattice degree of freedom, and the term is identical for all three layers $\ell=1,2,3$.
The results of the dispersion with this term are shown in Fig.~\ref{fig:bare_p2}.
This term breaks the exact $E(\vec{k})\leftrightarrow -E(-\vec{k})$ symmetry and produces a small gap at the Dirac points, larger than the remote hopping term,
but the overall magnitude of the effect is still small compared to the effect of the momentum-dependent tunneling term. Furthermore, the breaking of particle-hole symmetry appears to mostly shift the entire spectrum, which cannot explain the particle-hole asymmetric compressibility we observe.

The second set of terms arises from an inhomogeneous strain field. The correction to the Hamiltonian is:
\begin{equation}
\delta H_2^{\ell}(\vec{k}) = v_F \gamma \vec{\sigma}\cdot\vec{\mathcal{A}}^{\ell}(\vec{r}) + \frac{C_0}{2}[\vec{k}\cdot\vec{\mathcal{A}}^{\ell}(\vec{r})+\vec{\mathcal{A}}^{\ell}(\vec{r})\cdot\vec{k}]\sigma_0
+\frac{1}{2}[\vec{k}\cdot\vec{\xi}(\vec{r})+\vec{\xi}(\vec{r})\cdot\vec{k}],
\label{eq:H2}
\end{equation}
where $\mathcal{A}_x(\vec{r})=\partial_xu_x(\vec{r})-\partial_y u_y(\vec{r})$, $\mathcal{A}_y(\vec{r})=-[\partial_xu_y(\vec{r})-\partial_y u_x(\vec{r})]$, with $u_{i}(\vec{r})$ being the local displacement of the lattice at position $\vec{r}$ in direction $i\in x$, $y$ relative to unrelaxed structure. And $\xi_x(\vec{r}) = [(\frac{v_F}{2}+2D_0)\partial_x u_x(\vec{r})]\sigma_x + [(\frac{v_F}{2}+D_0)\partial_y u_x(\vec{r}) + D_0 \partial_x u_y(\vec{r})]\sigma_y$, $\xi_y(\vec{r}) = [(\frac{v_F}{2}+D_0)\partial_x u_y(\vec{r})+D_0\partial_y u_x(\vec{r})]\sigma_x + [(\frac{v_F}{2}+2D_0)\partial_y u_y(\vec{r})]\sigma_y$. 
Here, $v_F\gamma=-3.36$~eV, $C_0/a=0.94$~eV, and $D_0/a=-0.75$~eV. The first two terms in this Hamiltonian arise from the strain-induced pseudo-magnetic gauge field.

\begin{figure}[t!]
    \renewcommand{\thefigure}{S\arabic{figure}}
    \centering
    \includegraphics[width=0.8\columnwidth]{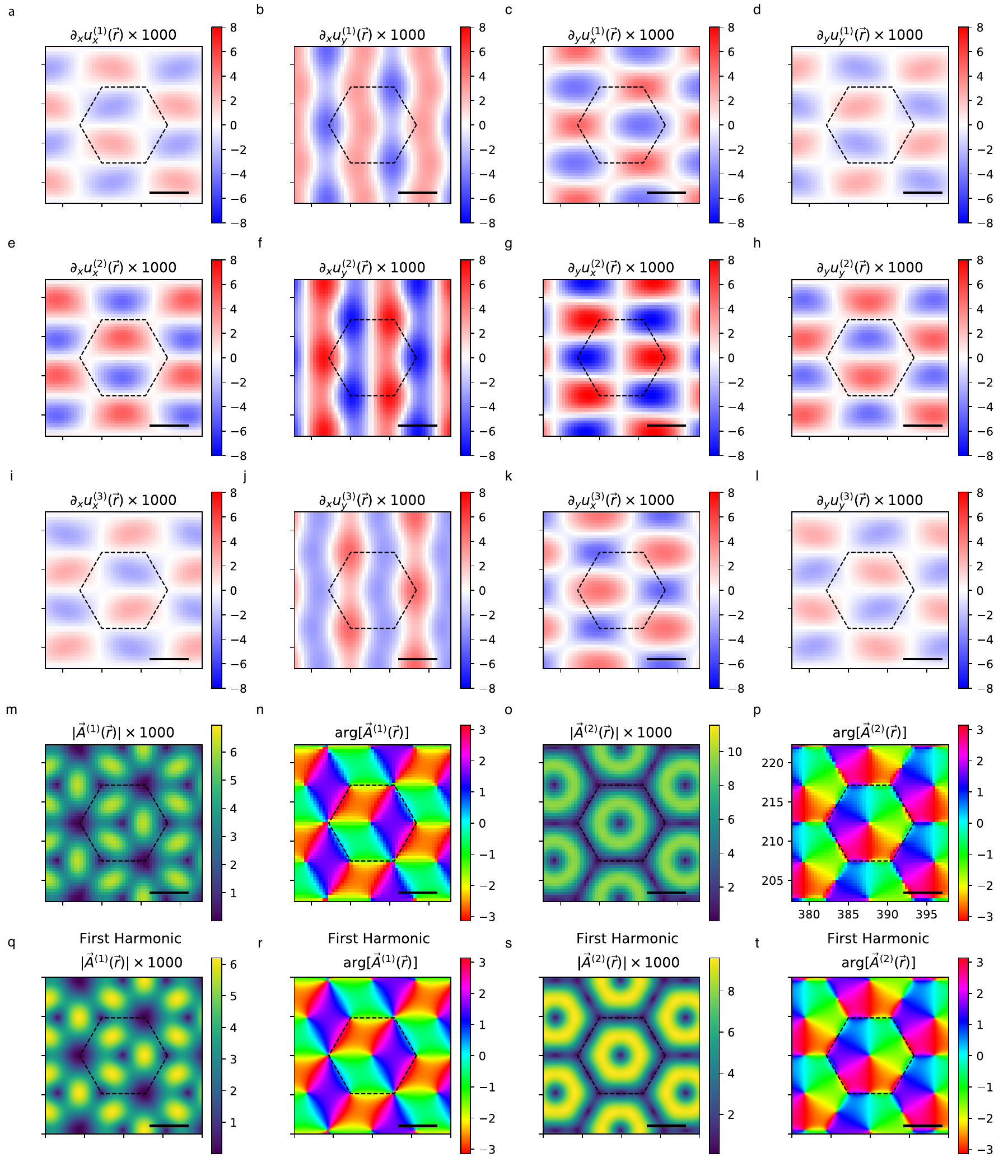}
    \caption{\textbf{The gradients of the strain field and the resulting pseudo-magnetic fields in the relaxed structure.} \textbf{a}-\textbf{d}, The gradients of the strain fields $\partial_i u_j(\vec{r})$ in the first layer. \textbf{e}-\textbf{h}, The gradients of the strain fields $\partial_i u_j(\vec{r})$ in the second layer. \textbf{i}-\textbf{l}, The gradients of the strain fields $\partial_i u_j(\vec{r})$ in the third layer. \textbf{m}(\textbf{n}), The amplitude (phase) of the pseudo-magnetic field on the first layer. \textbf{o}(\textbf{p}), The amplitude (phase) of the pseudo-magnetic field on the second layer. \textbf{q}(\textbf{r}), The amplitude (phase) of the first-harmonic of the pseudo-magnetic field on the first layer. \textbf{s}(\textbf{t}), The amplitude (phase) of the first-harmonic of the pseudo-magnetic field on the second layer. All scale bars are 5 nm.
} 
    \label{fig:strain_calculation}
\end{figure}

To characterize the strength and shape of these terms, we need to obtain the gradients $\partial_i u_j(\vec{r})$ of the strain fields in the relaxed structure. 
This can be calculated directly from the relaxed structure calculated in Sec.~3f using the configuration space method.
Figure~\ref{fig:strain_calculation}\textbf{a}-\textbf{l} shows a spatial map of all the components of $\partial_i u_j^{(\ell)}(\vec{r})$ at the center of a h-HTG domain, for each of the layers $\ell$.
These relaxation patterns satisfy the requisite $C_{3z}$ and $C_{2y}$ symmetry of the h-HTG domain.

o incorporate the effect of this strain field into the continuum model, we consider the first shell of its Fourier transform
\begin{equation}
\partial_i u_j (\vec{r})\approx c_{ij0}^{(\ell)} + \sum_{n=1}^6 c_{ijn}^{(\ell)}e^{i \vec{g}_n\cdot\vec{r}}
\end{equation}
where $\vec{g}_n = k_\theta(\cos\frac{2\pi (n-1)}{6},\sin\frac{2\pi(n-1)}{6})$.

We first consider the symmetry constraints on these coefficients. For the $c_{ij0}$ term, $C_{3z}$ symmetry enforces $c_{ij0}=a_0 \delta_{ij} + a_1 \epsilon_{ij}$, where $a_0$ and $a_1$ are real coefficients. We ignore $a_1$, which corresponds to a global rotation, but keep $a_0$. Next, we consider the $n=1,\dots,6$ coefficients $c_{ijn}^{(\ell)}$. By $C_{3z}$ and reality of $\partial_i u_j(\vec{r})$, all six coefficients can be obtained from $c_{ij1}^{(\ell)}$. Furthermore, since $\vec{g}_1 = k_\theta\hat{x}$, and the coefficients are obtained from derivatives along $i$, $c_{ij1} \sim (\vec{g}_1)_i$, we have that $c_{yj1}^{(\ell)}=0$. Thus, this leads to two non-zero complex parameters $c_{xx1}^{(\ell)}\equiv b_1^{(\ell)}$ and $c_{xy1}^{(\ell)}\equiv b_2^{(\ell)}$ per layer. Furthermore, $C_{2y}$ symmetry relates the parameters of the $\ell=1$ and $\ell=3$ layers. We can obtain a closed-form expression for the coefficients
\begin{equation}
\begin{split}
c_{ij0}^{(\ell)}&=a_0\delta_{ij}\\
c_{ijn}^{(\ell)}&=\begin{cases}
b_2^{(\ell)}\frac{(\vec{g}_n)_i(\vec{g}_n)_j}{k_\theta^2}
+ b_1^{(\ell)}\frac{(\vec{g}_n)_i\sum_{k}\varepsilon_{jk}(\vec{g}_n)_k}{k_\theta^2} \quad \text{if $n$ is odd,}\\
b_2^{(\ell)*}\frac{(\vec{g}_n)_i(\vec{g}_n)_j}{k_\theta^2}
+ b_1^{(\ell)*}\frac{(\vec{g}_n)_i\sum_{k}\varepsilon_{jk}(\vec{g}_n)_k}{k_\theta^2} \quad \text{if $n$ is even,}
\end{cases}
\end{split}
\end{equation}
where $\varepsilon_{ij}=(i\sigma_y)_{ij}$ is an anti-symmetric matrix.
From the relaxed structure shown in Fig.~\ref{fig:strain_calculation}, we extract the coefficients $a_0^{(1)}=a_0^{(3)}\approx -1\times 10^{-4}$, $a_0^{(2)}\approx 2\times 10^{-4}$,
$b_{1}^{(1)}=-b_1^{(3)*}\approx(-7.3+13.6i)\times 10^{-4}, b_1^{(2)}\approx(0-27.3i)\times 10^{-4}$, and all $b_2^{(\ell)}\approx 0\times 10^{-4}$. 
Here, we believe these coefficients are accurate down to $\approx 1\times 10^{-5}$ as they are obtained from numerical Fourier transformation for a finite window near the center of the domain (the relaxed h-HTG structure in the domain is not exactly periodic, even with lattice reconstruction). In Fig.~\ref{fig:strain_calculation}\textbf{q}-\textbf{t}, we show the pseudo-magnetic gauge fields calculated with the first-harmonic approximation following this analysis. This approximation shows good agreement with the full calculation of the strain pseudo-magnetic field (Fig.~\ref{fig:strain_calculation}\textbf{m}-\textbf{p}), indicating that this approximation is fairly accurate.

\begin{figure}[t!]
    \renewcommand{\thefigure}{S\arabic{figure}}
    \centering
    \includegraphics[width=0.4\columnwidth]{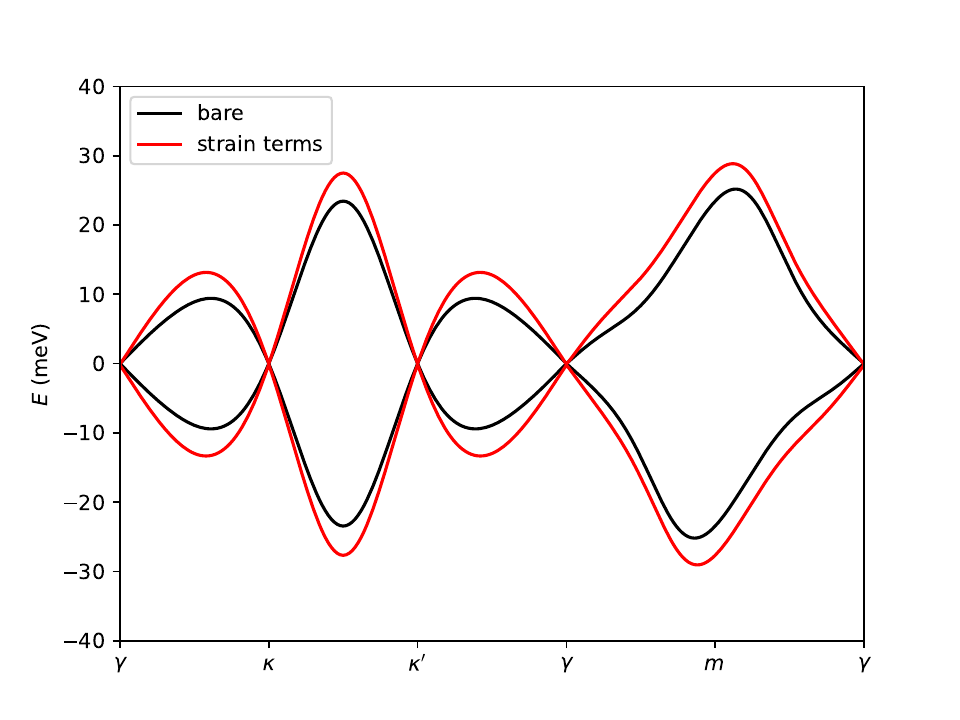}
    \caption{\textbf{Effect of the gradients of the strain field.} Zoom-in of the band structure near the CNP with the gradients of the strain field (red) and with the bare model (black). These terms do not break the particle-hole symmetry, nor gap out the Dirac points in the band structure.
} 
    \label{fig:bare_strain}
\end{figure}

Figure~\ref{fig:bare_strain} shows the band structure with the gradients of the strain field taken from the above analysis. The main effect of all the terms in $\delta H_2$ is to increase the bandwidth of the flat bands. In fact, these terms do not break particle-hole symmetry nor gap out the Dirac points.

The result of this analysis is that momentum-dependent tunneling has by far the biggest effect in introducing particle-hole asymmetry and gapping out the Dirac node at $\gamma$. Neither remote interlayer hopping nor intralayer moiré terms can reproduce our data.

\subsection{d. Displacement field dependence}

\begin{figure}[t!]
    \renewcommand{\thefigure}{S\arabic{figure}}
    \centering
    \includegraphics[width=.96\columnwidth]{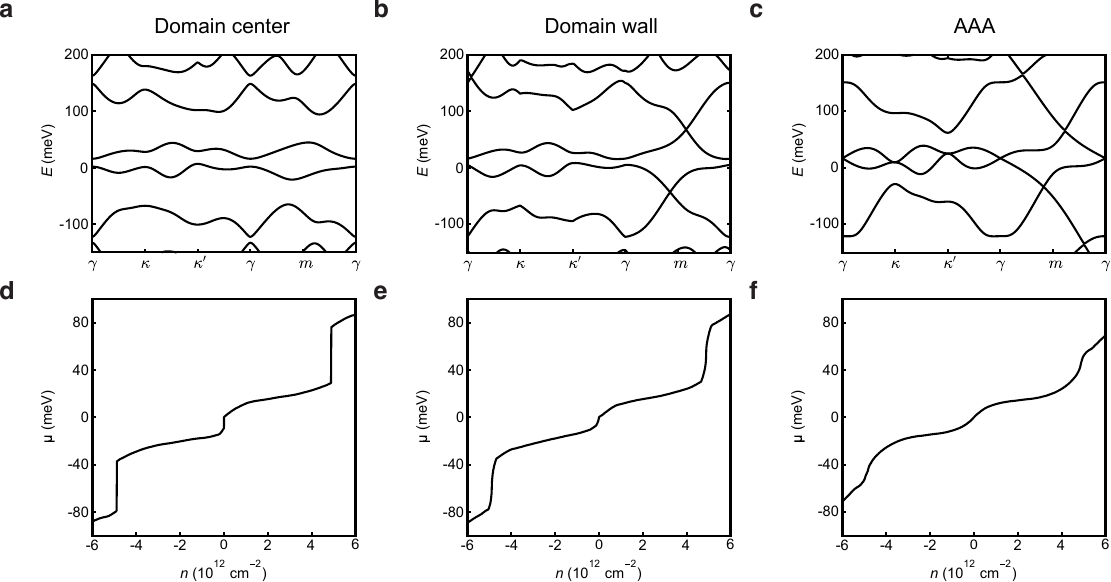}
    \caption{\textbf{Band structure and chemical potential evolution at $D=0.45$~V/nm.} \textbf{a}-\textbf{c}, Band structure for $D=0.45$~V/nm at a domain center (\textbf{a}), a domain wall (\textbf{b}), and an AAA site (\textbf{c}). \textbf{d}-\textbf{f}, Corresponding $\mu(n)$ for each respective supermoiré location. 
} 
    \label{fig:finite_D_band}
\end{figure}

Gating the sample with the bottom gate introduces a nonzero $D$ that roughly linearly depends on $n$~\cite{mccann2006asymmetry,foutty2023tunable}. We therefore also calculate the band structure at the domain center, domain wall, and AAA site at $D=0.45$ V/nm (Fig.~\ref{fig:finite_D_band}\textbf{a}-\textbf{c}), which is the estimated value in our device~\cite{foutty2023tunable} corresponding to the carrier density at moiré filling factor $\nu=4$ for $\theta=1.45^{\circ}$ (neglecting corrections from screening related to charge redistribution between layers). Comparing the band structure at $D=0$ and $D=0.45$ V/nm, we see that the domain center remains gapped at $\nu=\pm4$, while the domain wall and AAA site remain gapless. Quantitatively, at the domain center, the $\nu=\pm4$ superlattice gaps are smaller at finite $D$. The corresponding $\mu(n)$ shows the largest increase at $\nu=4$ ($n\approx5\times10^{12}~\rm{cm}^{-2}$) in the domain center, and the smallest increase on AAA sites (Fig.~\ref{fig:finite_D_band}\textbf{d}-\textbf{f}), consistent with our observations. These band structures and chemical potential calculations were used to produce the theoretical simulation in Fig.~3\textbf{h}-\textbf{i} and Fig.~\ref{fig:sim_eg} below. Experimentally, the effective displacement field can be tuned over a narrow (10s of mV/nm) window by changing the voltage difference between the SET tip and the sample~\cite{foutty2024mapping}. We do not observe a difference in behavior over the accessible range.

\subsection{e. Hofstadter spectrum calculations}

In this section, we describe the calculation of the Hofstadter spectrum of HTG.  
The procedure follows closely that described in Ref.~\cite{uri2023superconductivity}, with the addition of the momentum-dependent tunneling term. The detailed derivation is included below.

Our starting point is the continuum model Hamiltonian in Eq.~\ref{eq:BMHam}.  
First, we parameterize the moiré reciprocal lattice vectors $\bm{G}_{\rm{M}}$ as 
\begin{equation}
\bm{G}_{\rm{M}} \equiv \bm{G}_{\rm{M,12}}=\bm{G}_{\rm{M,23}}=g\begin{pmatrix}\frac{1}{2} & -1 \\ \frac{\sqrt{3}}{2} & 0\end{pmatrix}
=
\begin{pmatrix}
\Delta_x & -2 \Delta_x \\ \Delta_y & 0
\end{pmatrix}
\end{equation}
where $g=\frac{4\pi\sin\theta}{\sqrt{3}a}$, $\Delta_x\equiv g/2$, $\Delta_y\equiv \sqrt{3}g/2$, and $\bm{G}_{\rm{M},\it{l'l}}\equiv \bm{G}_{l'}-\bm{G}_{l}$.

A perpendicular magnetic field $B$ is incorporated via minimal substitution of the momentum relative to some reference momentum.  We take the reference momentum to be the average of all the $K$-points (which is simply the $\vec{K}_2$ point in this system), $\vec{K}_{\rm{av}}=\frac{1}{3}\sum_l \vec{K}_l=\vec{K}_2$.
We also define $\bm{G}_{\rm{av}}=\frac{1}{3}\sum_l\bm{G}_l$ for future use.

Incorporating a magnetic field via minimal substitution sends $k_x\rightarrow \pi_x = k_x + B y/2$ and $k_y\rightarrow \pi_y=k_y-B x/2$.
These operators satisfy the commutation relation $[\pi_x,\pi_y]=iB$.  
We further define $X=k_x-By/2$ and $Y=k_y+Bx/2$, which commute with $\pi_{x,y}$ and satisfy $[X,Y]=-iB$.

The intralayer kinetic Hamiltonian for layer $l$ is given by
\begin{equation}
H_l = ve^{i\theta_l}\sigma^+[\pi_x-i\pi_y-({K}_{l,x}-iK_{l,y}-{K}_{\rm{av},x}+iK_{\rm{av},y})] + h.c. \equiv  v \sqrt{2B}e^{i\theta_l}\sigma^+(b^\dagger-S_l) + h.c.
\end{equation}
where $b^\dagger\equiv \frac{1}{\sqrt{2B}}[\pi_x-i\pi_y]$ is the Landau level raising operator satisfying $[b,b^\dagger]=1$, $\sigma^+=\left(\begin{smallmatrix}0&1\\0&0\end{smallmatrix}\right)$ is a raising operator acting on sublattice space,
and $S_l\equiv (K_{l,x}-i K_{l,y}-K_{\rm{av},x}+iK_{\rm{av},y})/\sqrt{2B}$.

The interlayer hopping term requires additional care. 
First, notice that $T_{l',l}(\vec{r},\vec{k})$ in Eq.~\ref{eq:tunnDef} contains both $\vec{r}$ and $\vec{k}$, so naively would depend on the ordering of $e^{-i(\vec{Q}^j_{l',l})\cdot\vec{r}}$ and $w^j_{l',l}(\vec{k})$.  
However, the ordering turns out to be irrelevant since these two terms commute: this must be the case since the tunneling process conserves total absolute momentum (this is straightforward for the $j=0$ term, for which $\vec{Q}^0_{l',l}=0$, and must be true for the $j\neq 0$ terms as well since they are related by the microscopic $C_{3z}$ symmetry).
In order to perform minimal substitution, we must obtain an analytic expression for the tunneling in terms of $\vec{k}$.
We first express the momentum-dependent hopping amplitude in Eq.~\ref{eq:wk} as
\begin{equation}
\begin{split}
    w^j_{l-1,l}(\vec{k}) &= w_0 + \frac{\rm{d}\omega}{\rm{d}\it{k}} (|\vec{k}+ \vec{K}_{\rm{av}} + \bm{G}_{\rm{av}} \vec{m}_j + (\bm{G}_{l}-\bm{G}_{\rm{av}})\vec{m}_j| - K_D )\\
    &\approx w_0 + \frac{\rm{d}\omega}{\rm{d}\it{k}} \left([\vec{k}+ (\bm{G}_{l}-\bm{G}_{\rm{av}})\vec{m}_j]\cdot\frac{\vec{K}_{\rm{av}}+\bm{G}_{\rm{av}}\vec{m}_j}{|\vec{K}_{\rm{av}}+\bm{G}_{\rm{av}}\vec{m}_j|} + |\vec{K}_{\rm{av}}+\bm{G}_{\rm{av}}\vec{m}_j| - K_D \right)
    \end{split}
\end{equation}
which can then be minimally substituted $\vec{k}\rightarrow\vec{\pi}$ and the matrix elements in the Landau level basis can be straightforwardly calculated.
Next, the matrix elements of the $e^{-i \vec{q}\cdot\vec{r}}$ terms are calculated by expanding
\begin{equation}
e^{-i\vec{q}\cdot\vec{r}}=e^{\frac{i}{B}[q_x Y - q_y X]}e^{\frac{i}{B}[-q_x \pi_y + q_y \pi_x]}
\end{equation}
into a product of two factors acting on the Hilbert spaces of $(X,Y)$ and $(\pi_x,\pi_y)$ separately.
The vectors $\vec{q}$ that appear in the Hamiltonian are of the form $\vec{q}=\vec{Q}^j_{l',l}=\bm{G}_{\rm{M},\it{l'l}}\vec{m}_j=d_x \Delta_x \hat{x} + d_y \Delta_y \hat{y}$ where $d_x$ and $d_y$ are integers.
Thus, the term $e^{\frac{i}{B}[q_x Y - q_y X]}=e^{\frac{i}{B}[d_x \Delta_x Y - d_y \Delta_y X]}$
acting on an eigenstate $\ket{X}$ of $X$ results in $e^{\frac{i}{B}[q_x Y - q_y X]}\ket{X}=e^{\frac{i}{B}(-q_y(X + \frac{d_x}{2}\Delta_x))}\ket{X+d_x \Delta_x}$.
For particular magnetic fields $B=p\Delta_x\Delta_y/(2\pi q)$ with positive integers $p,q$, the Hamiltonian becomes periodic with respect to a shift $\ket{X}\rightarrow\ket{X+p\Delta_x}$.
We can therefore Fourier transform along the $X$ direction, labeling states as
\begin{equation}
\ket{J;x_0,k_0}\propto \sum_{m=-\infty}^{\infty}e^{\frac{2\pi i k_0}{p}(x_0+J+mp)} \ket{X=(x_0+J+mp)\Delta_x}
\end{equation}
in which $x_0,k_0\in [0,1]$ are good quantum numbers of $H$, $J=0,1,\dots, p-1$, and we define $\ket{J+p;x_0,k_0}=\ket{J;x_0,k_0}$.
Then, 
\begin{equation}
e^{\frac{i}{B}(q_x Y - q_y X)}\ket{J;x_0,k_0}=e^{-2\pi i(\frac{d_y q}{p} x_0 + \frac{d_x}{p} k_0)}e^{-\frac{2\pi iq d_y}{p}(J+d_x/2)}\ket{J+d_x;x_0,k_0}
\end{equation}
For the other factor, $e^{\frac{i}{B}[-q_x \pi_y + q_y \pi_x]}$, we work in the Landau level basis $\ket{n}$ satisfying $b^\dagger b \ket{n}=n \ket{n}$, $n=0,1,\dots$.  
The matrix elements are 
$\bra{n^\prime}e^{\frac{i}{B}(-q_x \pi_y + q_y \pi_x)}\ket{n}=D_{n^\prime n}([q_x+iq_y]/\sqrt{2B})$ where
\begin{equation}
D_{n^\prime n}(z)=\bra{n^\prime}e^{z b^\dagger - z^* b}\ket{n}=\begin{cases}
e^{-\frac{|z|^2}{2}} (z)^{n^\prime-n} \sqrt{\frac{n!}{n^\prime !}}L_{n}^{(n^\prime-n)}(|z|^2) & \mathrm{if}\;\;\;\; n^\prime \geq n \\
e^{-\frac{|z|^2}{2}} (-z^*)^{n-n^\prime} \sqrt{\frac{n^\prime!}{n !}}L_{n^\prime}^{(n-n^\prime)}(|z|^2) & \mathrm{else} ,
\end{cases}
\end{equation}
where $L_{n}^{(\alpha)} $ are the generalized Laguerre polynomials.
Putting everything together, we can express the interlayer tunneling matrix elements in the basis $\ket{n,J,\sigma,l;x_0,k_0}$.
We define the unit vector $\vec{K}_{\mathrm{unit},j}\equiv \frac{\vec{K}_{\mathrm{av}}+\bm{G}_{\mathrm{av}}\vec{m}_j}{|\vec{K}_{\mathrm{av}}+\bm{G}_{\mathrm{av}}\vec{m}_j|}$ and ``offset'' vector $\vec{k}_{\mathrm{off},j,l}=(\bm{G}_l-\bm{G}_{\rm{av}})\vec{m}_j$.
The interlayer hopping matrix elements from layer $l$ to $l^\prime=l\pm 1$, with a relative interlayer shift $\vec{d}_0=\vec{d}_{b(t)}$, is then given by
\begin{equation}
\begin{split}
\bra{n^\prime,J^\prime,\sigma^\prime,l^\prime;x_0,k_0}
&T_{l^\prime l}(\vec{r}-\vec{d}_0,\vec{\pi})
\ket{n,J,\sigma,l;x_0,k_0}
=\\
\sum_{j=0}^2& 
e^{i (\vec{Q}^{j}_{l^\prime l})\cdot \vec{d}_0}
(T_j)_{\sigma^\prime \sigma}
e^{-2\pi i(\frac{q}{p}d_y^{jl^\prime l} x_0 + \frac{1}{p}d_x^{jl^\prime l} k_0)}e^{-\frac{2\pi iq}{p} d_y^{jl^\prime l}(J+d_x^{jl^\prime l}/2)}
\delta_{J^\prime,J+d_x^{j l^\prime l}}
\\
&\times \left[D_{n^\prime n}\left(\frac{{Q}^j_{l^\prime,l,x}+iQ^j_{l^\prime,l,y}}{\sqrt{2B}}\right)\left(w_0 + \frac{\mathrm{d}\omega}{\mathrm{d}\it{k}}
(\vec{k}_{\mathrm{off},j,l}\cdot\vec{K}_{\mathrm{unit},j}+|\vec{K}_{\mathrm{av}}+\bm{G}_{\mathrm{av}}\vec{m}_j|-K_D)\right)
\right. \\
&\;\;\;\;\;\; + D_{n^\prime,n-1}\left(\frac{{Q}^j_{l^\prime,l,x}+iQ^j_{l^\prime,l,y}}{\sqrt{2B}}\right)\frac{\mathrm{d}\omega}{\mathrm{d}\it{k}}\sqrt{\frac{B n}{2}}(K_{\mathrm{unit},j,x}-iK_{\mathrm{unit},j,y})\\
&\left.\;\;\;\;\;\; + D_{n^\prime,n+1}\left(\frac{{Q}^j_{l^\prime,l,x}+iQ^j_{l^\prime,l,y}}{\sqrt{2B}}\right)\frac{\mathrm{d}\omega}{\mathrm{d}\it{k}}\sqrt{\frac{B (n+1)}{2}}(K_{\mathrm{unit},x}+iK_{\mathrm{unit},j,y})\right]
\end{split}
\end{equation}
where the terms involving $D_{n^\prime,n\pm 1}$ come from the Landau level raising and lowering operators present in the momentum-dependent tunneling $w^j_{l^\prime,l}(\vec{\pi})$, and we have defined the integers $d_{x(y)}^{jl^\prime l}=Q_{l^\prime l,x(y)}^j/\Delta_{x(y)}$.

For each $(x_0,k_0)$, we follow this procedure to construct the Hamiltonian matrix which can be diagonalized to obtain a set of energies $\{E^{x_0 k_0}_i\}$.
The total density of states is then obtained by
\begin{equation}
\mathrm{DOS}(E)=\frac{2}{2q A_{\mathrm{M}}}\left\langle\sum_{i}\delta(E-E^{x_0k_0}_i)\right\rangle_{x_0 k_0}
\end{equation}
where the average is taken over $(x_0,k_0)$ pairs, and $A_{\mathrm{M}}=4\pi^2/\mathrm{det}(\bm{G}_{\mathrm{M}})$ is the moir\'e unit cell area.
The factor of 2 in the numerator accounts for the spin degeneracy.

Because this system is not $C_{2z}$ symmetric, a separate calculation must be carried out for the $K^\prime$ valley.
To obtain the $K^\prime$ valley, we exploit the fact that the $C_{2z}$ operation relates the $K^\prime$ valley of the $h$ domain of HTG with the $K$ valley of the $\bar{h}$ domain.  The spectrum of the $K^\prime$ valley can therefore be obtained by repeating the calculation above but for the shift $\vec{d}_{t(b)}$ corresponding to the opposite domain.

We sweep magnetic fields, between $B=0.5$ T and $B=11$ T, using $1 \leq p, q \leq 40$.
The full Hamiltonian matrix can then be constructed for a fixed $x_0,k_0\in [0,1]$, with the Landau level index truncated to a maximum value $n<N_{\mathrm{max}}$, that is large enough to ensure convergence (additional care must be taken in constructing the Hamiltonian matrix to ensure that the finite cutoff does not contribute a state at low energy). For a magnetic field $B=p\Delta_x\Delta_y/(2\pi q)$, we use $N_{\mathrm{max}}=\lfloor 40 q/p \rfloor $, where $\lfloor ... \rfloor$ is the floor function. 
The spectrum is then sampled over $(x_0,k_0)$ chosen randomly from a uniform distribution, and we take at least $100$ samples.

\begin{figure}[t!]
    \renewcommand{\thefigure}{S\arabic{figure}}
    \centering
    \includegraphics[width=0.9\columnwidth]{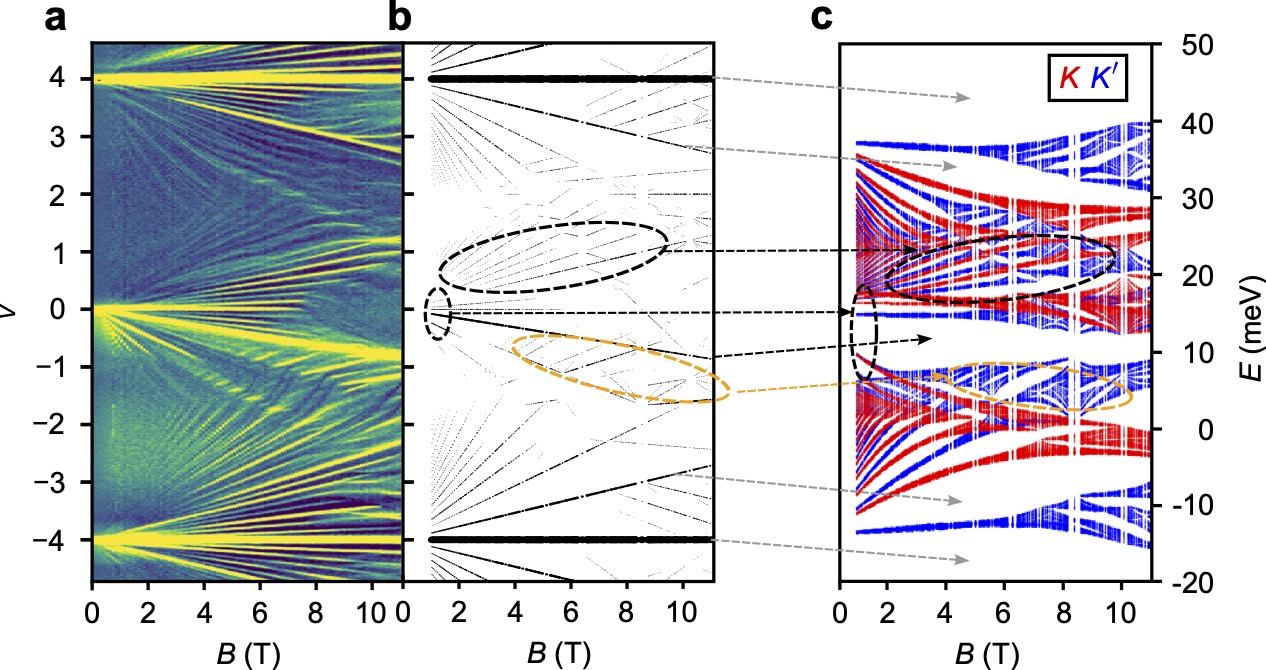}
    \caption{\textbf{Mapping between the Landau fan and Hofstadter spectrum.} \textbf{a}, d$\mu$/d$n$ as a function of $\nu$ and $B$, reproduced from Fig.~4a. \textbf{b}-\textbf{c}, Plot of Landau fan gaps (\textbf{b}) and the Hofstadter spectrum (\textbf{c}), respectively reproduced from Fig.~4c and Fig.~ED7. Gaps with different zero-field intercepts are indicated with dashed arrows with different colors (moiré filling factor $\nu = 0$, black; $\nu = -2$, orange;  $\nu = \pm4$, gray). Select landau levels from $\nu = 0$ and Chern insulators emanating from $\nu = -2$ are encircled by black and orange ovals, respectively. Note that in the Landau fan calculation (\textbf{b}) the size of each dot reflects the gap magnitude, while in the Hofstadter spectrum calculation (\textbf{c}) the density of dots reflects the density of states. 
} 
    \label{fig:hoft_comparisons}
\end{figure}

The resulting Hofstadter spectra for valleys $K$ and $K^\prime$ in the $h$ domain are shown in Fig.~ED7. The Hofstadter spectra are different for each valley, which reflects the fact that $C_{2z}$ symmetry is broken within a single domain. Below, we describe in more detail the mapping between experimental observations and qualitative features in each theoretical representation. For ease of comparison, we reproduce Fig.~4\textbf{a},\textbf{c} and Fig.~ED7 in Fig.~\ref{fig:hoft_comparisons}. The sizes of the dots in Fig.~\ref{fig:hoft_comparisons}\textbf{b} represent the magnitude of the corresponding gaps, whereas the density of dots in the Hofstadter spectrum in Fig.~\ref{fig:hoft_comparisons}\textbf{c} reflects the density of states. Therefore, a one-to-one connection can be made between gaps in Fig.~\ref{fig:hoft_comparisons}\textbf{b} and the white spaces (representing zero density of states) in Fig.~\ref{fig:hoft_comparisons}\textbf{c}. 

To facilitate comparison, we highlight with dashed arrows the correspondence between a subset of the gaps in the calculated Landau fan and in the Hofstadter spectrum (Fig.~\ref{fig:hoft_comparisons}\textbf{b}-\textbf{c}). The color of the arrows reflects the zero field intercept of each gap (black for $\nu=0$, gray for $\nu=\pm4$, and orange for $\nu=-2$). The zero field intercept of each gap is related to its location relative to the van Hove singularities of the Hofstadter spectrum in each valley. Specifically, each gap will stem from an empty ($s=-2$), half-filled ($s=0$), or filled ($s=2$) state in each valley at zero field, where $s$ is the zero-field intercept of a single valley. The zero-field intercept of a given gapped state is therefore $\nu=s_{K}+s_{K'}$, which can be $\pm4$, $\pm2$, or $0$. Note that a state with $|\nu|=2$ indicates that one valley has $s=0$ while the other has $|s|=2$. Therefore, the observation of states with zero-field intercepts of $\nu=2$ in the experiment provides strong evidence for broken valley symmetry due to local $C_{2z}$ symmetry breaking within the HTG domains.

We emphasize that a gap only appears in regions where the Hofstadter spectra of both valleys are simultaneously gapped. This is why we do not see prominent gaps at high field near the charge neutrality point, where the two valleys have overlapping spectra. Similarly, broadening from temperature and/or disorder likely precludes resolution of the mini-gaps between subbands at low magnetic fields, except for the most prominent states near the charge neutrality point and the superlattice gaps (back and grey arrows, respectively). Regions with a large gap between subbands in one valley correspond to swaths were mini-gaps from the other valley are visible (e.g. within the orange oval in Fig.~\ref{fig:hoft_comparisons}). This explains why there are finite windows dominated by a series of gaps of a given $\nu$ as filling and magnetic field are tuned.


\section{2. Ruling out a secondary moiré period and large twist angle mismatch}

\begin{figure}[t!]
    \renewcommand{\thefigure}{S\arabic{figure}}
    \centering
    \includegraphics[width=0.45\columnwidth]{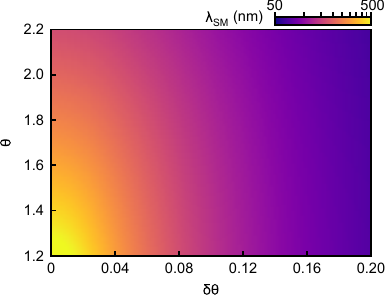}
    \caption{\textbf{Dependence of the supermoiré wavelength on interlayer twist angles.} Supermoiré wavelength $\lambda_\mathrm{SM}$ as a function of twist angle $\theta$ and twist angle mismatch $\delta \theta$ at zero strain. Here we assume the twist angle between graphene layers 1 and 2 is $\theta_{12} = \theta + \delta \theta/2$, while for layers 2 and 3, $\theta_{23} = \theta - \delta \theta/2$. The curve in Fig.~2\textbf{g} corresponds to a horizontal line cut at $\theta = 1.45^\circ$.
    }  
    \label{fig:twist_angle_mismatch}
\end{figure}


Here, we discuss several pieces of experimental evidence that indicate the absence of a second moiré periodicity that would arise from a large twist angle mismatch. First, optical images of the HTG sample during the fabrication process (Fig.~ED10) indicate that the three graphene layers are closely aligned with each other: $\theta_{12} \approx \theta_{23} \lesssim 2^\circ$. This suggests the interlayer angles are similar. 

Second, we observe no signatures of additional superlattice features in either transport or local compressibility measurements. During transport characterization, we measured up to a carrier density $n = 7 \times 10^{12}$ cm$^{-2}$, which precludes a second moiré periodicity corresponding to a second twist angle less than 1.74$^\circ$. Likewise, no additional superlattice gaps are visible in local electronic compressibility measurements across the entire sample (Fig.~ED2) nor are there moiré-of-moiré features~\cite{ren2024tunable,xie2024strong}, indicating that even locally, there is not a secondary moiré period corresponding to an angle below $1.57^{\circ}$. In fact, the large supermoiré domains that we observe suggest that the two interlayer angles are likely similar. In the presence of a small twist angle mismatch, the supermoiré wavelength $\lambda_\mathrm{SM}$ rapidly decreases with increasing twist angle mismatch $\delta \theta$ (Fig.~\ref{fig:twist_angle_mismatch}):
\begin{equation}
\lambda_\mathrm{SM}(\theta, \delta\theta) = \frac{a}{2\sqrt{1-2\rm{cos}(\theta) \rm{cos}\left(\delta\theta/2\right)+\rm{cos}^2(\theta)}}.\label{eq:lambdaSM}
\end{equation}
Geometrical considerations show that a device with one large interlayer twist angle will generally have a small supermoiré wavelength, and realistic values of strain ($|\epsilon|<0.7\%$ ~\cite{xie2019spectroscopic,kerelsky2019maximized,choi2019electronic,yoo2019atomic,mcgilly2020visualization,kerelsky2021moireless,turkel2022orderly_disorder,nuckolls2023quantum,kim2023imaging}) cannot produce the large supermoiré wavelength we observe (see Supplementary Sec. 3).
We remark that Eq.~\ref{eq:lambdaSM} for the supermoir\'e period is valid for small twist angles, and that it does not imply the existence of an exactly commensurate structure at the atomic scale.

Furthermore, a large interlayer twist would result in decoupled dispersive, Dirac-like bands. In a magnetic field, this would produce outward-sloping moiré superlattice gaps due to Chern numbers from Landau levels of Dirac-like bands~\cite{hoke2024uncovering, kim2022evidence, shen2023dirac, pierce2024tunable, park2021tunable, hao2021electric}. In contrast, we observe $C = 0$ superlattice gaps at $\nu = \pm 4$ independent of magnetic field (Fig.~4\textbf{a}), demonstrating that such Dirac bands are not present. A moderate twist angle mismatch is further excluded because our observations are qualitatively distinct from those of Ref.~\cite{ren2024tunable}, where despite a similar range of interlayer angles ($\theta_{12}=1.64^{\circ}, \theta_{23}=1.33^{\circ}$), two sets of superlattice features are present, indicating that the system does not locally relax even for relatively small mismatch $\delta\theta=0.31^{\circ}$.

Finally, theoretical calculations of the band structure and Hofstadter spectrum of HTG match well with the experiment (Fig.~4, Fig.~\ref{fig:D0_zoomin}, Fig.~\ref{fig:hoft_comparisons}). Such close agreement would not be expected if the two interlayer angles were very different and there were two independent moiré wavelengths. We therefore conclude that the most likely scenario is that the sample starts with relatively similar interlayer angles and has relaxed into large domains with a single moiré periodicity.


\section{3. Effect of strain on the HTG supermoiré domain and moiré structure}
Here we elaborate on the expected behavior when at least one of the graphene layers is (globally) strained, taking $\theta=1.45^{\circ}$ and $\delta\theta=0$ for simplicity, and discuss its impact on moiré and supermoiré structures. We denote the three reciprocal wavevectors of layer $l$ in the absence of strain as ${\vec{G}}_{l}^{i}$, where $|{\vec{G}}_{l}^{i}|=4\pi/\sqrt{3}a$,
and $i=1$, 2, or 3, with ${\vec{G}}_{l}^{3}=-({\vec{G}}_{l}^{1}+{\vec{G}}_{l}^{2})$ defined for convenience in mathematical notation. For a relative twist angle $\theta_{ll'}$ between the layers $l$ and $l'$, their reciprocal wavevectors are related by
\begin{equation}
{\vec{G}}_{l}^{i}=
\begin{pmatrix}
\mathrm{cos}(\theta_{ll'}) & -\mathrm{sin}(\theta_{ll'})\\
\mathrm{sin}(\theta_{ll'}) & \mathrm{cos}(\theta_{ll'})
\end{pmatrix}
{\vec{G}}_{l'}^{i}.
\end{equation}
We separately discuss in detail below unaxial strain, biaxial strain, shear strain, and their combination, all of which can affect the supermoiré periodicity. Lastly, we show how strain affects moiré and supermoiré periodicity differently, causing a separation of moiré and supermoiré length scales in the system.

\subsection{a. Biaxial strain}

\begin{figure}[b!]
    \renewcommand{\thefigure}{S\arabic{figure}}
    \centering
    \includegraphics[width=1\columnwidth]{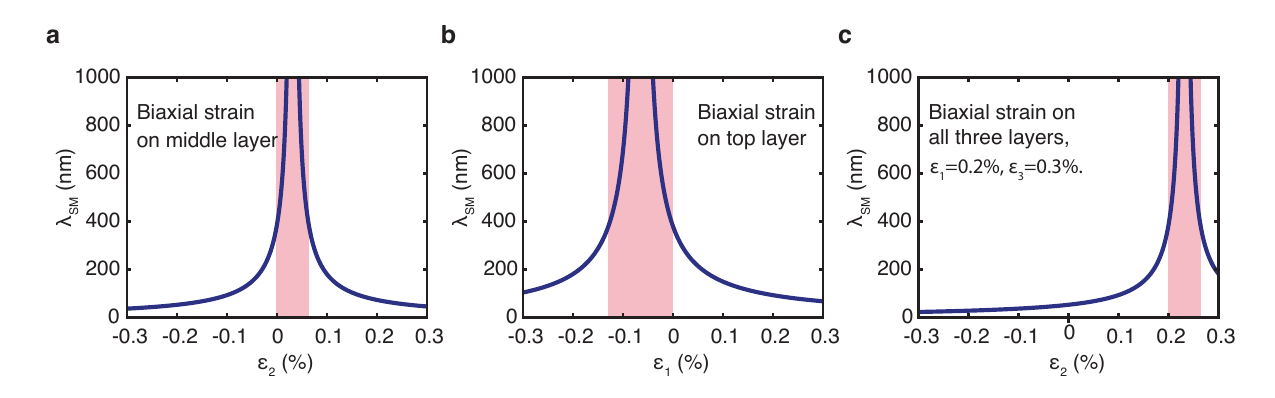}
    \caption{\textbf{Effect of biaxial strain on $\lambda_{\mathrm{SM}}$.} \textbf{a}, $\lambda_{\mathrm{SM}}$ as a function of biaxial heterostrain $\epsilon_2$ applied only on the middle layer. \textbf{b}, $\lambda_{\mathrm{SM}}$ as a function of biaxial heterostrain $\epsilon_1$ applied only on the top layer. \textbf{c}, Dependence of $\lambda_{\mathrm{SM}}$ when isotropic biaxial stain is applied to all three layers, with $\epsilon_1=0.2\%$ and $\epsilon_3=0.3\%$ fixed. Here $\lambda_{\mathrm{SM}}^1=\lambda_{\mathrm{SM}}^2=\lambda_{\mathrm{SM}}^3\equiv\lambda_{\mathrm{SM}}$ for all cases because the strain is biaxial and isotropic. The range where $\lambda_{\mathrm{SM}}$ is enhanced compared to the zero strain scenario is shaded in pink.} 
    \label{fig:biaxial_strain}
\end{figure}

We first consider isotropic biaxial strain, because anisotropic biaxial strain can be decomposed into a combination of uniaxial strain and isotropic biaxial strain. If the biaxial strain on layer $l$ is $\epsilon_l$ (positive $\epsilon_l$ corresponds to stretching), then its reciprocal lattice vector is $\vec{G}^{\it{'i}}_{\it{l}}=\frac{1}{1+\epsilon_l}\vec{G}^{\it{i}}_{\it{l}}$. Then the supermoiré reciprocal lattice vector can be derived by 
\begin{equation}
\vec{G}^{\it{i}}_{\rm{SM}}=\vec{G}^{\it{'i}}_{1}+\vec{G}^{\it{'i}}_{3}-2\vec{G}^{\it{'i}}_{2},
\end{equation}
and the supermoiré periodicity is 
\begin{equation}
\lambda_{\mathrm{SM}}=\frac{4\pi}{\sqrt{3}|\vec{G}^{\it{i}}_{\rm{SM}}|}.
\end{equation}
Here we obtain the same supermoiré periodicity $\lambda_{\mathrm{SM}}$ along different directions because isotropic biaxial strain preserves the 6-fold rotational symmetry of graphene.

In general, isotropic biaxial strain drastically changes the supermoiré domain size. We plot several representative configurations in Fig.~\ref{fig:biaxial_strain}, including strain on a single graphene layer and strains of different magnitudes on all layers. For each case, the supermoiré periodicity diverges when the condition $\epsilon_2=1-\frac{2(1-\epsilon_1)(1-\epsilon_3)\rm{cos}(\theta)}{2-\epsilon_1-\epsilon_3}$ is met. This implies that the entire HTG sample would preferentially relax into a single moiré-periodic domain at this point. Away from the divergent point, the supermoiré periodicity monotonically decreases as a function of strain. The pink shading in Fig.~\ref{fig:biaxial_strain} indicates the range over which the supermoiré wavelength is enhanced compared to the zero strain case. We conclude that biaxial strain can contribute to the increased supermoiré unit cell area compared to the unstrained theory prediction (Fig.~2\textbf{c}), and to its further enhancement after thermal cycling (Fig.~2\textbf{f}). The relatively symmetric supermoiré domains in the second round of measurements can be explained by biaxial strain. However, to obtain the anisotropic supermoiré pattern in the first round of measurements, uniaxial and/or shear strain must also be present.

\begin{figure}[t]
    \renewcommand{\thefigure}{S\arabic{figure}}
    \centering
    \includegraphics[width=.45\columnwidth]{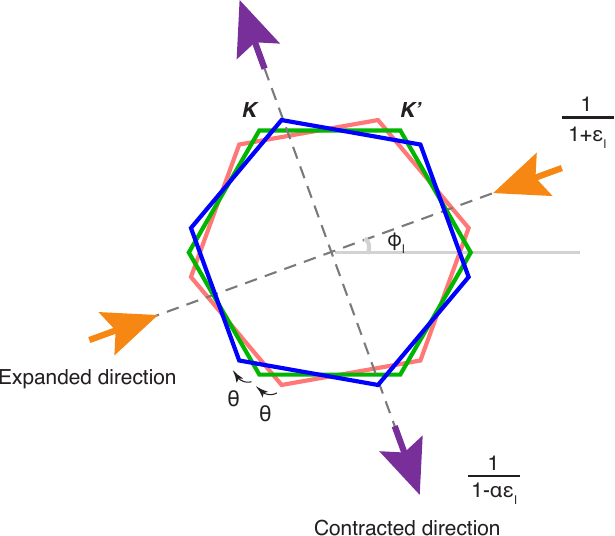}
    \caption{\textbf{Schematic of uniaxial strain in reciprocal space.} The reciprocal lattice of the three graphene layers with equal interlayer twist angles $\theta$ are represented in blue, green, and red, respectively. The uniaxial strain experienced by layer $l$ of graphene is denoted by $\epsilon_l$, along a direction of $\phi_l$ relative to the reciprocal lattice of the middle layer in $k$-space. For graphene, the lattice constant is then deformed by $\alpha=0.16\epsilon_l$ perpendicular to the strain direction.   
} 
    \label{fig:uniaxial_strain_config}
\end{figure}

\subsection{b. Uniaxial strain}
We next discuss the impact of uniaxial strain. For uniaxial strain, we denote the strain of layer $l$ by $\epsilon_l$, along a direction with angle $\phi_l$ relative to the reciprocal lattice of the middle layer in $k$-space (Fig.~\ref{fig:uniaxial_strain_config}). Here $\epsilon_l>0$ ($\epsilon_l<0$) again indicates the expansion (contraction) of the graphene lattice in the strained direction. The lattice is also contracted (expanded) in the perpendicular direction by $\alpha\epsilon_l$, where $\alpha=0.16$ is the Poisson ratio of graphene~\cite{kerelsky2019maximized}. Following the deformation of the reciprocal lattice vectors, we can obtain the supermoiré periodicity in all three directions. The corresponding mathematical relations are given below.

When layer $l$ experiences uniaxial strain $\epsilon_l$ along the direction $\phi_l$, its reciprocal lattice along direction $i$ is deformed to
\begin{equation}
\vec{G}^{\it{'i}}_{\it{l}}=
\begin{pmatrix}
\mathrm{cos}(\phi_{l}) & \mathrm{sin}(\phi_{l})\\
-\mathrm{sin}(\phi_{l}) & \mathrm{cos}(\phi_{l})
\end{pmatrix}
\begin{pmatrix}
\frac{1}{1+\epsilon_l} & 0\\
0 & \frac{1}{1-\alpha\epsilon_l}
\end{pmatrix}
\begin{pmatrix}
\mathrm{cos}(\phi_{l}) & -\mathrm{sin}(\phi_{l})\\
\mathrm{sin}(\phi_{l}) & \mathrm{cos}(\phi_{l})
\end{pmatrix}
\vec{G}^{\it{i}}_{\it{l}}
\end{equation}
After deformation, the $i$-th reciprocal wavevector corresponding to the moiré pattern between layer $l$ and $l'$ is
\begin{equation}
\vec{G}^{\it{i}}_{\rm{M},\it{ll'}}=\vec{G}^{\it{'i}}_{\it{l}}-\vec{G}^{\it{'i}}_{\it{l'}}.
\end{equation}
Then the $i$-th supermoiré reciprocal lattice wavevector is
\begin{equation}
\vec{G}^{\it{i}}_{\rm{SM}}=\vec{G}^{\it{i}}_{\rm{M,12}}-\vec{G}^{\it{i}}_{\rm{M,23}}.
\end{equation}
Thus the supermoiré periodicities in real space are 
\begin{equation}
\lambda_{\mathrm{SM}}^1=\left|\frac{2\pi{\hat{Q}}\vec{G}^{2}_{\rm{SM}}}{\vec{G}^{1}_{\rm{SM}}\cdot\hat{Q}\vec{G}^{2}_{\rm{SM}}}\right|,~~~\lambda_{\mathrm{SM}}^2=\left|\frac{2\pi{\hat{Q}}\vec{G}^{1}_{\rm{SM}}}{\vec{G}^{2}_{\rm{SM}}\cdot\hat{Q}\vec{G}^{1}_{\rm{SM}}}\right|,~~~\lambda_{\mathrm{SM}}^3=|\vec{\lambda}_{\rm{SM}}^1+\vec{\lambda}_{\rm{SM}}^2|,
\end{equation}
where $\hat{Q}$ is an operator that rotates a vector 90$^\circ$ counter-clockwise. 

\begin{figure}[t!]
    \renewcommand{\thefigure}{S\arabic{figure}}
    \centering
    \includegraphics[width=1\columnwidth]{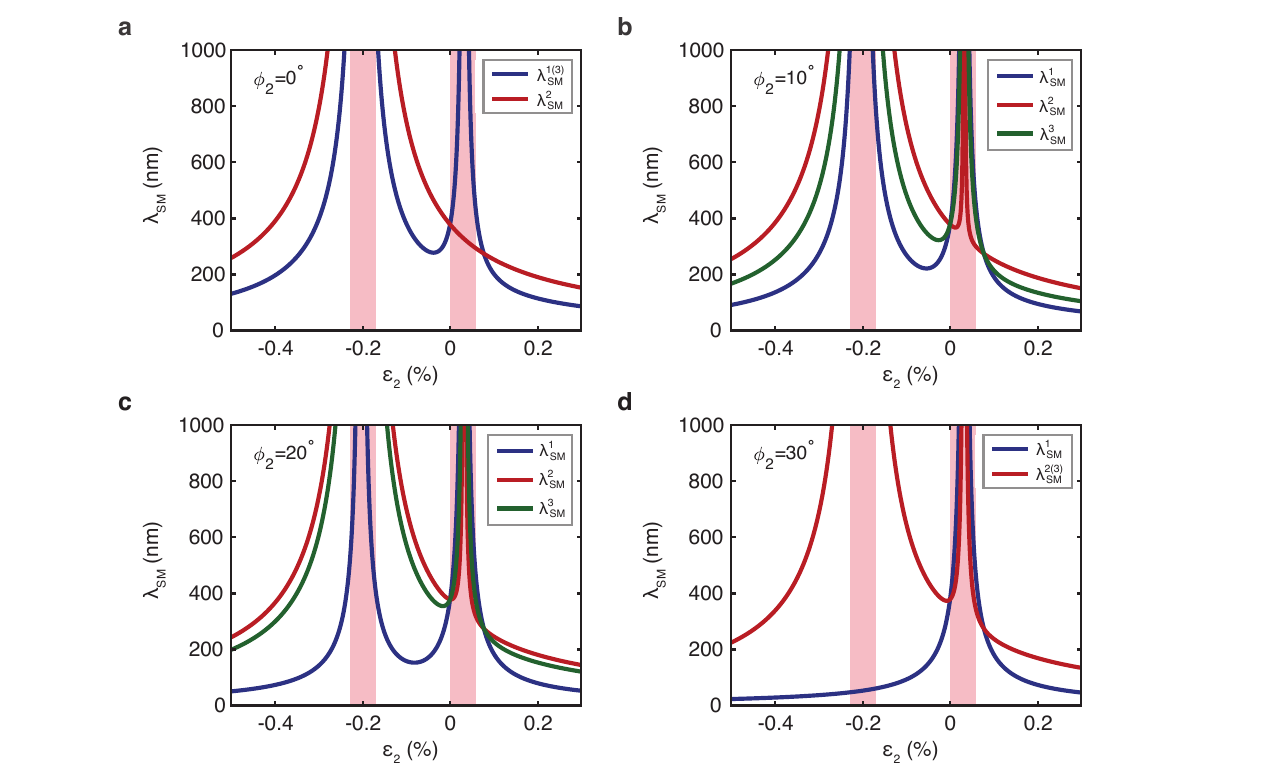}
    \caption{\textbf{Dependence of the supermoiré wavelength on uniaxial strain applied to the middle layer.} \textbf{a}-\textbf{d}, Supermoiré wavelengths as a function of uniaxial strain $\epsilon_2$ on the middle layer along a direction $\phi_2=0^\circ$ (\textbf{a}), $\phi_2=10^\circ$ (\textbf{b}), $\phi_2=20^\circ$ (\textbf{c}), and $\phi_2=30^\circ$ (\textbf{d}). Blue, green, and red colors respectivly indicate the supermoiré wavelength $\lambda_{\mathrm{SM}}^i$ along directions $i=1$, 2, and 3. For uniaxial strain along high symmetry directions, we have $\lambda_{\mathrm{SM}}^1=\lambda_{\mathrm{SM}}^3$ in \textbf{a}, and $\lambda_{\mathrm{SM}}^2=\lambda_{\mathrm{SM}}^3$ in \textbf{d}. The ranges over which the supermoiré unit cell area is enhanced compared to the zero strain scenario are shaded in pink.} 
    \label{fig:uniaxial_strain}
\end{figure}

We start from the simplest case where only one layer experiences uniaxial strain. We first consider a uniaxial strain on the middle layer ($\epsilon_2\neq0$, $\epsilon_1=\epsilon_3=0$) along a direction $\phi_2$. We limit our discussion to $0^\circ\le\phi_2\le30^\circ$, which can be related by symmetry to other angles. 
We then plot the supermoiré periodicity along all three directions, $\lambda_{\mathrm{SM}}^1$, $\lambda_{\mathrm{SM}}^2$, and $\lambda_{\mathrm{SM}}^3$, as a function of $\epsilon_2$ at four representative $\phi_2$ angles: 0$^\circ$, 10$^\circ$, 20$^\circ$, and 30$^\circ$, in Fig.~\ref{fig:uniaxial_strain}\textbf{a}-\textbf{d}. If the strain lies along a high-symmetry direction, two out of the three supermoiré periodicities are equal. This leads to $\lambda_{\mathrm{SM}}^2=\lambda_{\mathrm{SM}}^3$ when $\phi_2=0^\circ$ in Fig.~\ref{fig:uniaxial_strain}\textbf{a}, and $\lambda_{\mathrm{SM}}^1=\lambda_{\mathrm{SM}}^3$ when $\phi_2=30^\circ$ in Fig.~\ref{fig:uniaxial_strain}\textbf{d}. 

When $\epsilon_2=0$, the three supermoiré wavelengths are equal ($\lambda_{\mathrm{SM}}=384$~nm when $\theta=1.45^\circ$). As the uniaxial strain on the second layer increases, the supermoiré periodicity along a particular direction can diverge at very small strain $\epsilon_2=1-\rm{cos}(\theta)\approx0.032\%$ (for example, $\lambda_{\mathrm{SM}}^{1(3)}$ in Fig.~\ref{fig:uniaxial_strain}\textbf{a}), at some intermediate strain strength $\epsilon_2=-[1-\rm{cos}(\theta)]/\alpha\approx-0.2\%$ (for example, for all directions in Fig.~\ref{fig:uniaxial_strain}\textbf{a}), or both. When the uniaxial strain on the second layer is close to one of these values, the size of the supermoiré unit cell diverges, and can be significantly enhanced beyond the size predicted by theory at zero strain (pink shaded regions). We note that even for cases where the supermoiré wavelength $\lambda_{\mathrm{SM}}^i$ is larger along all three directions than its value in the absence of strain, the supermoiré unit cell area can still be smaller than that in the absence of strain.
As the strain deviates from a point of divergence, the supermoiré periodicity rapidly decreases. In particular, for $\epsilon_2>0.032\%$ and $\epsilon_2<-0.2\%$, the supermoiré periodicity decreases in all three directions, which quickly renders the supermoiré unit cell area much smaller than the theoretically predicted value at zero strain. Within the empirical strain range in graphene moiré systems~\cite{xie2019spectroscopic,kerelsky2019maximized,choi2019electronic,yoo2019atomic,mcgilly2020visualization,kerelsky2021moireless,turkel2022orderly_disorder,nuckolls2023quantum,kim2023imaging}, the size of the supermoiré unit cell changes non-monotonically as a function of the strain, which can explain the various supermoiré areas we observe in Fig.~2.

\begin{figure}[t!]
    \renewcommand{\thefigure}{S\arabic{figure}}
    \centering
    \includegraphics[width=1\columnwidth]{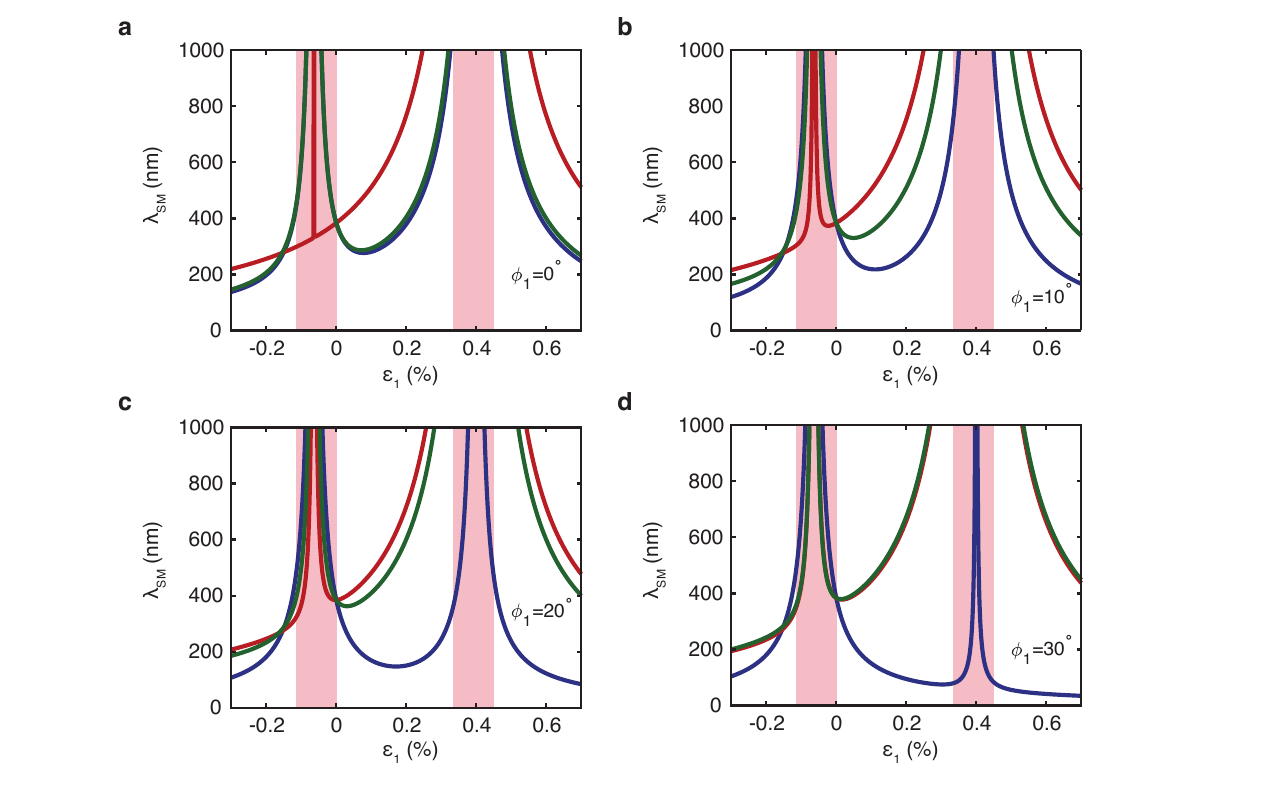}
    \caption{\textbf{Dependence of the supermoiré wavelength on uniaxial strain applied to the top layer.} \textbf{a}-\textbf{d}, Supermoiré wavelengths as a function of uniaxial strain $\epsilon_1$ on the top layer along a direction $\phi_1=0^\circ$ (\textbf{a}), $\phi_1=10^\circ$ (\textbf{b}), $\phi_1=20^\circ$ (\textbf{c}), and $\phi_1=30^\circ$ (\textbf{d}). Blue, green, and red colors respectivly indicate the supermoiré wavelength $\lambda_{\mathrm{SM}}^i$ along directions $i=1$, 2, and 3. The ranges over which the supermoiré unit cell area is enhanced compared to the zero strain scenario are shaded in pink.}
    \label{fig:uniaxial_strain_top_layer}
\end{figure}

We next discuss the case where only the top or bottom layer experiences uniaxial strain. Without loss of generality, we assume the uniaxial strain is on the top layer ($\epsilon_1\neq0$) along the $\phi_1$ direction ($0^{\circ}\le\phi_1\le30^{\circ}$; again, other angles can be mapped back to this range by symmetry). The corresponding supermoiré wavelengths along each direction are plotted for various angles in Fig.~\ref{fig:uniaxial_strain_top_layer}\textbf{a}-\textbf{d}. We observe that at two different strain strengths $\epsilon_1=-2[1-\rm{cos}(\theta)]=-0.064\%$ and $\epsilon_1=2[1-\rm{cos}(\theta)]/\alpha=0.4\%$, all three supermoiré wavelengths $\lambda_{\mathrm{SM}}$ concurrently diverge. 
Within a reasonable range of strain parameters, the supermoiré unit cell area can either increase or decrease relative to the theoretically predicted value without strain.

\subsection{c. Shear strain}
Shear strain can also change the supermoiré structure. Note that in general any shear deformation of a single layer of graphene can be decomposed into pure shear and rotation in the layer angle. The latter was discussed in Supplementary Section 2. Therefore, we only consider pure shear in graphene~\cite{cocco2010gap,engelke2023topological}. We assume the $l$-th layer has shear strain $\gamma_l$ along the direction with angle $\phi_l$. Then the reciprocal lattice of the $l$-th layer along direction $i$ is deformed to
\begin{equation}
\vec{G}^{\it{'i}}_{\it{l}}=
\begin{pmatrix}
\mathrm{cos}(\phi_{l}) & \mathrm{sin}(\phi_{l})\\
-\mathrm{sin}(\phi_{l}) & \mathrm{cos}(\phi_{l})
\end{pmatrix}
\begin{pmatrix}
1 & -\gamma_l\\
-\gamma_l & 1
\end{pmatrix}
\begin{pmatrix}
\mathrm{cos}(\phi_{l}) & -\mathrm{sin}(\phi_{l})\\
\mathrm{sin}(\phi_{l}) & \mathrm{cos}(\phi_{l})
\end{pmatrix}
\vec{G}^{\it{i}}_{\it{l}}
\end{equation}
The resulting supermoiré reciprocal lattice is then $\vec{G}^{\it{i}}_{\rm{SM}}=\vec{G}^{\it{'i}}_{1}+\vec{G}^{\it{'i}}_{3}-2\vec{G}^{\it{'i}}_{2}$ and the corresponding supermoiré wavelength can be calculated.

\begin{figure}[t!]
    \renewcommand{\thefigure}{S\arabic{figure}}
    \centering
    \includegraphics[width=1\columnwidth]{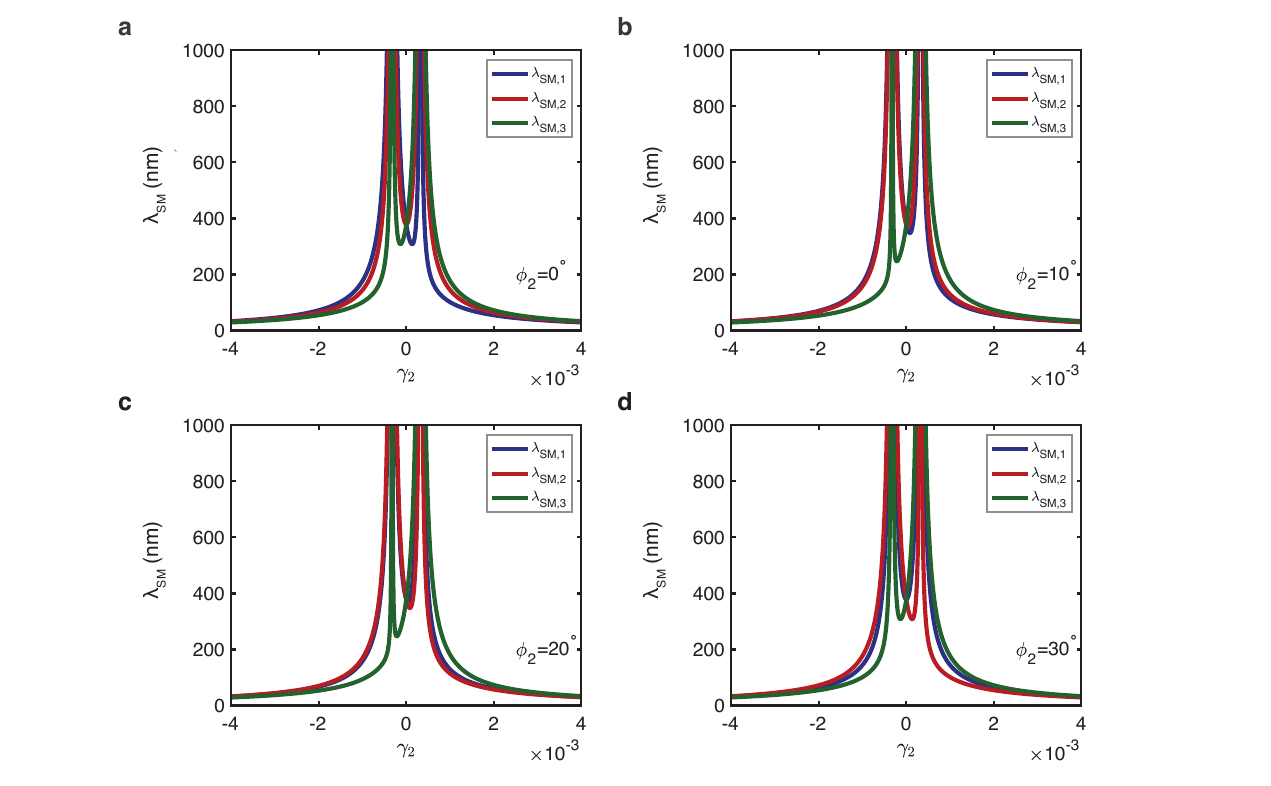}
    \caption{\textbf{Dependence of the supermoiré wavelength on shear strain applied to the middle layer.} \textbf{a}-\textbf{d}, Supermoiré wavelengths as a function of shear strain $\gamma_2$ on the middle layer along a direction $\phi_2=0^\circ$ (\textbf{a}), $\phi_2=10^\circ$ (\textbf{b}), $\phi_2=20^\circ$ (\textbf{c}), and $\phi_2=30^\circ$ (\textbf{d}). Blue, green, and red colors respectively indicate the supermoiré wavelength $\lambda_{\mathrm{SM}}^i$ along directions $i=1$, 2, and 3. } 
    \label{fig:shear_strain_middle_layer}
\end{figure}

We here only consider a representative scenario where only one graphene layer experiences shear strain. In Fig. \ref{fig:shear_strain_middle_layer}, we plot the supermoiré lengths in the presence of shear strain on the middle layer only; panel \textbf{a} to \textbf{d} correspond to different shear strain directions. The calculations show that the supermoiré wavelength becomes anisotropic. Shear strain can either enhance or shrink supermoiré domain size.

\subsection{d. Strain parameters that produce the experimental results}
\begin{figure}[t!]
    \renewcommand{\thefigure}{S\arabic{figure}}
    \centering
    \includegraphics[width=1\columnwidth]{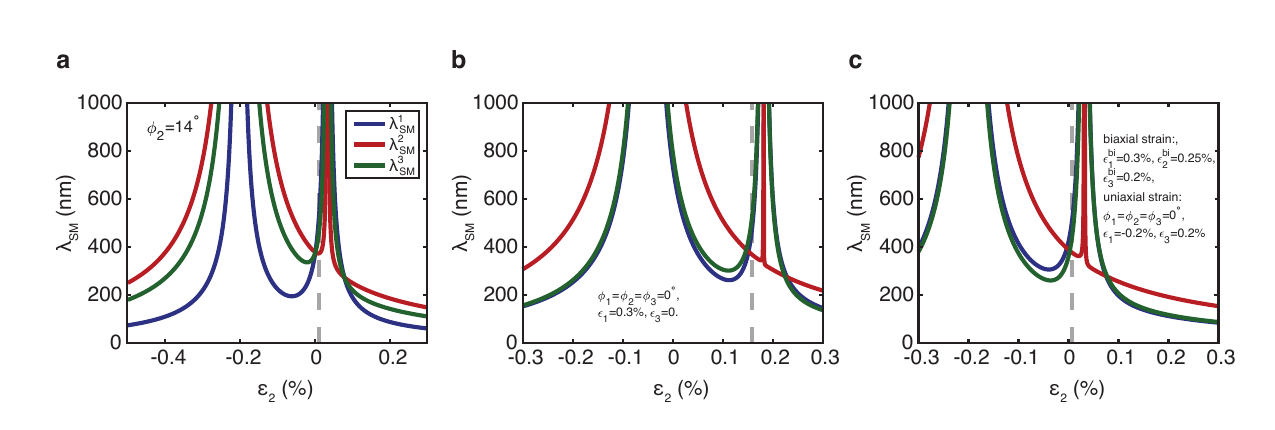}
    \caption{\textbf{Example strain configurations that reproduce the supermoiré periodicity in experiment.} \textbf{a}, There is only uniaxial strain on the middle layer along $\phi_2=14^\circ$. At $\epsilon_2=0.009\%$ (gray dashed line), the supermoiré periodicity in Fig.~3\textbf{a} is obtained. \textbf{b}, There is uniaxial strain on the top and middle layers along $\phi_1=\phi_2=0^\circ$ direction. When $\epsilon_1=0.3\%$, and $\epsilon_2=0.158\%$ (gray dashed line), the supermoiré periodicity in Fig.~3\textbf{a} is obtained. \textbf{c}, Both uniaxial strain and biaxial strain are present in the system. The isotropic biaxial strain on the three graphene layers are $\epsilon_1^{\rm{bi}}=0.3\%$, $\epsilon_2^{\rm{bi}}=0.25\%$, and $\epsilon_3^{\rm{bi}}=0.2\%$. And the uniaxial strain $\epsilon_1=-0.2\%$, $\epsilon_1=0.042\%$, $\epsilon_3=0.2\%$, all along the $\phi=0^\circ$ direction (gray dashed line), reproduce the supermoiré periodicity in Fig.~3\textbf{a}.} 
    \label{fig:possible_configurations}
\end{figure}

Realistically, it is likely that more than one layer is strained, making the parameter space for calculating the supermoiré periodicity very large. A similar analysis involving both uniaxial strain and biaxial strain as above can be applied to calculate the supermoiré periodicity more generally. We highlight a few representative cases that can reproduce the supermoiré periodicity shown in Fig.~3\textbf{a} in the main text in Fig.~\ref{fig:possible_configurations} to demonstrate the range of strain parameters that can match the experimental observations; the value of strain that matches the experimental signal is denoted by gray dashed lines in each panel. 

Note that multiple strain configurations can reproduce the experimentally observed supermoiré periodicities. These also span different magnitudes of strain. We therefore cannot deduce quantitative conclusions about the specific strain configuration nor the overall magnitude of strain. However, the analysis does provide several important implications. First, the enhanced domain sizes we observe, including their anisotropy, can be explained by strain. We are unaware of other mechanisms that can increase domain size, and therefore attribute our observations to strain. Second, while multiple strain configurations can enhance supermoiré domain size, this corresponds to a small fraction of the overall available parameter space; most strain configurations would decrease domain area. Our observations therefore suggest that strain in a given device may tend to stabilize at configurations with enhanced domain sizes for energetic reasons. Furthermore, generic strain configurations often involve divergence of the supermoiré wavelength along two or three directions. The former case would produce stripe-like reconstruction rather than the relatively symmetric triangular domains that we observe.

\subsection{e. Separation of the moiré and supermoiré length scale by strain}
\begin{figure}[t!]
    \renewcommand{\thefigure}{S\arabic{figure}}
    \centering
    \includegraphics[width=1\columnwidth]{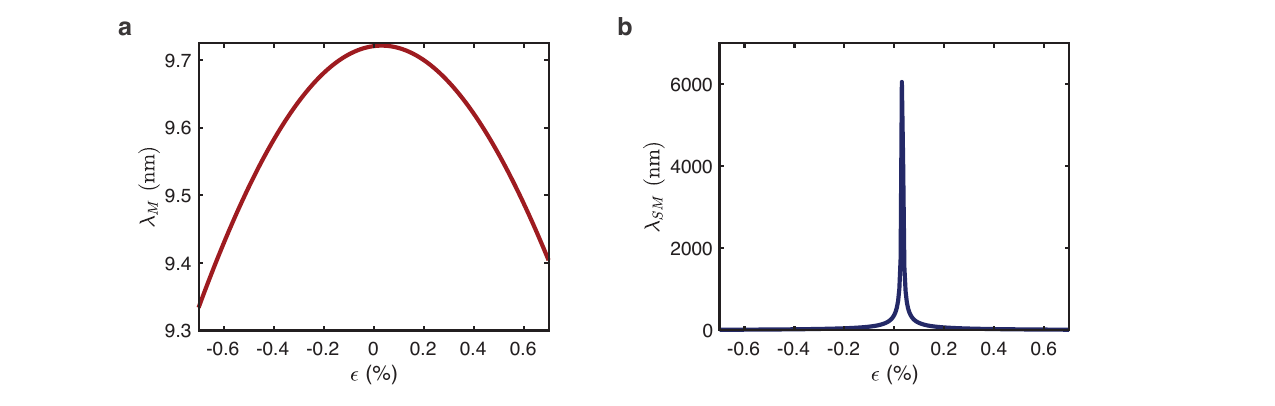}
    \caption{\textbf{Effect of strain on moiré and supermoiré length scales in HTG.} \textbf{a}, The dependence of the moiré length $\lambda_{\rm{M}}$ on isotropic biaxial strain $\epsilon$ applied to the middle layer of HTG with interlayer twist angles $\theta=1.45^\circ$. The change in $\lambda_{\rm{M}}$ is less than 4\% for the experimentally relevant strain range $|\epsilon|<0.7\%$. \textbf{b}, Dependence of the supermoiré length $\lambda_{\rm{SM}}$ on strain, showing a divergence near $\epsilon=0.032\%$.}
    \label{fig:seperation_of_length}
\end{figure}

Here we demonstrate that, with a moderate amount of strain, the moiré wavelength of the system remains almost unchanged, but the supermoiré length can diverge. This allows for a separation of the two length scales in HTG. We consider a representative case where the top and bottom graphene layers are not strained, but the middle graphene layer experiences isotropic biaxial strain $\epsilon$, where $\epsilon>0$ ($\epsilon<0$) means the graphene lattice is expanded (contracted). Similar considerations also apply for other strain configurations. Under this configuration, the reciprocal lattice vector of the middle layer is changed to $|G_2'|=|G_2|/(1+\epsilon)$. Therefore, the moiré reciprocal vector between adjacent layers is 
$\vec{G}_{M,12}=\vec{G}_1-\vec{G}_2'$, where
\begin{equation}
\vec{G}_1=
\begin{pmatrix}
\mathrm{cos}(\theta) & -\mathrm{sin}(\theta)\\
\mathrm{sin}(\theta) & \mathrm{cos}(\theta)
\end{pmatrix}
\vec{G}_2
\end{equation}
The resulting moiré wavelength between adjacent layers is then $\lambda_{\mathrm{M},12}=2\pi/(\sqrt{3}/2|G_{\mathrm{M},12}|)=a/\sqrt{1+(\frac{1}{1+\epsilon})^2-\frac{2\cos(\theta)}{1+\epsilon}}$. In a reasonable range of strain based on prior works in twisted graphene systems ($|\epsilon|<0.7\%$~\cite{xie2019spectroscopic,kerelsky2019maximized,choi2019electronic,yoo2019atomic,mcgilly2020visualization,kerelsky2021moireless,turkel2022orderly_disorder,nuckolls2023quantum,kim2023imaging}), and for our experimental twist angle, $\lambda_{\mathrm{M}}$ is minimally affected by the strain over the relevant range of parameters (less than 4\% change at maximum), as shown in Fig.~\ref{fig:seperation_of_length}\textbf{a}.

This contrasts with the effect of strain on the supermoiré period, which can be significantly more pronounced. For the supermoiré wavelength in this scenario, $\vec{G}_\mathrm{SM}=\vec{G}_1+\vec{G}_3-2\vec{G}_2'$, which leads to $\lambda_{\mathrm{SM}}=\frac{a}{2|\frac{1}{1+\epsilon}-\cos(\theta)|}$. The supermoiré wavelength diverges when $\frac{1}{1+\epsilon}=\cos(\theta)$ and can thus vary substantially as a function of strain for realistic parameter ranges (Fig.~\ref{fig:seperation_of_length}\textbf{b}). This strain tunability of the supermoiré wavelength is plotted (over a narrower range) in Fig. 2h in the main text.

\subsection{f. Energetics of the supermoiré domain structure under strain}

We calculate the lattice relaxation under strain to further support the conclusion that we can tune the supermoiré domains without changing the local electronic properties. Here we use the “configuration space” approach in Refs.~\cite{carr2018relaxation} and \cite{zhu2020modeling} to obtain the relaxed lattice structure under strain when the total elastic and stacking energy is minimized. Since this method works in configuration space rather than real space, it can describe the relaxed structures of incommensurate patterns or, in this case, structures with very large supermoiré periods. We use the model parameters reported in Ref.~\cite{carr2018relaxation} and show the results below.

We consider the case of global biaxial strain on the middle layer, since we expect it to lead to the most dramatic effects in terms of separation of moiré and supermoiré length scales, but the calculation can be applied to any type of strain.

In this approach, the total elastic energy is taken to consist of intralayer and interlayer terms.  The main modification to the algorithm beyond Ref.~\cite{carr2018relaxation} and \cite{zhu2020modeling} is to incorporate global strain into the intralayer energy. Focusing on just a single layer, the intralayer elastic energy expressed in real space takes the form
\begin{align}
E=\sum_{ijkl}\int d^2\vec{r}\epsilon_{ij}(\vec{r})C_{ijkl}\epsilon_{kl}(\vec{r}),
\end{align}
where $C_{ijkl}$ is the elastic tensor, and $\epsilon_{ij}(\vec{r})=\frac{1}{2}(\partial_i u_j+\partial_j u_i)$ is the strain tensor, and $u_i(\vec{r})$ is the local displacement of the lattice at position $\vec{r}$ in direction $i\in{x,y}$ relative to unrelaxed structure. The total elastic and stacking energy is then minimized with respect to finite functions $u_i(\vec{r})$ that depends only on the local atomic configurations at $\vec{r}$. To incorporate the global biaxial strain, we take $u_i(\vec{r})=u_{0,i}(\vec{r})+\tilde{u}_i(\vec{r})$, where $u_{0,x}(\vec{r})=\epsilon r_x$; $u_{0,y}(\vec{r})=\epsilon r_y$ represent a global biaxial strain by amount and $\tilde{u}_i(\vec{r})$ encodes the local displacements on top of this background.  This results in the strain tensor $\epsilon_{ij}(\vec{r})=\epsilon\delta_{ij}+\tilde{\epsilon}_{ij}(\vec{r})$, where $\tilde{\epsilon}_{ij}(\vec{r})=\frac{1}{2}(\partial_i\tilde{u}_j(\vec{r})+\partial_j\tilde{u}_i(\vec{r}))$.

This gives intralayer energy 
\begin{align}
E=\sum_{ijkl}\int d^2\vec{r}(\epsilon\delta_{ij}+\tilde{\epsilon}_{ij}(\vec{r}))C_{ijkl}(\epsilon\delta_{kl}+\tilde{\epsilon}_{kl}(\vec{r}))=\epsilon^2A\sum_{ij}C_{iijj}+\sum_{ijkl}\int d^2\vec{r}\tilde{\epsilon}_{ij}(\vec{r})C_{ijkl}\tilde{\epsilon}_{kl}(\vec{r}),
\end{align}
where $A$ is the total sample area. Compared to the zero-strain case, other than a constant energy offset $\epsilon^2A\sum_{ij}C_{iijj}$, the intralayer elastic energy is of the exact same mathematical form, but with $\tilde{u}_{i}(\vec{r})$ instead of $u_i(\vec{r})$. Thus, we can model the lattice relaxation under strain with the same method as in Ref.~\cite{devakul2023magic}, simply by scaling the lattice constant of the middle graphene layer by a factor of $1+\epsilon$.

\begin{figure}[t!]
    \renewcommand{\thefigure}{S\arabic{figure}}
    \centering
    \includegraphics[width=1\columnwidth]{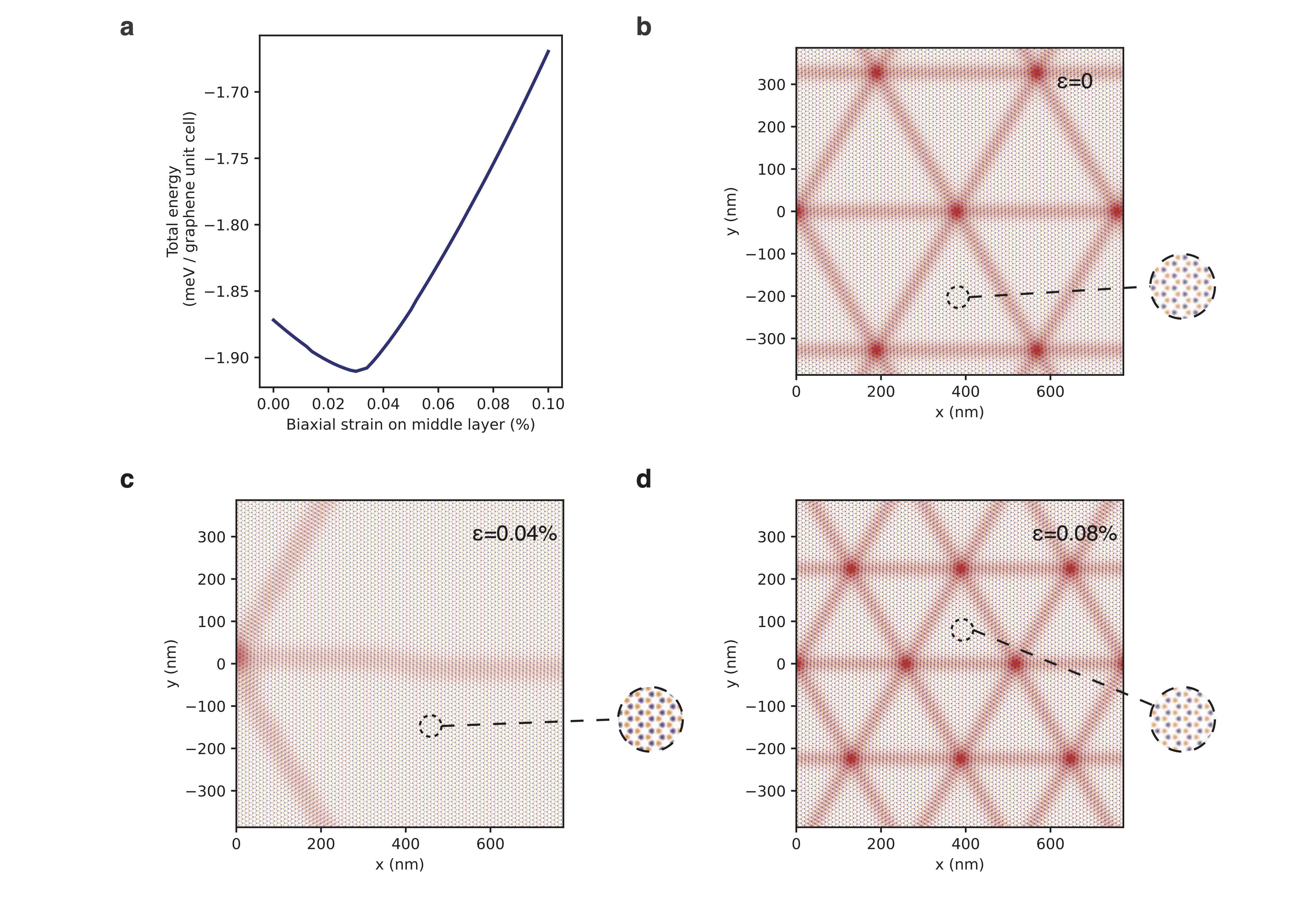}
    \caption{\textbf{a}, The total energy (stacking energy + elastic energy) of the relaxed lattice configuration at different biaxial strain strength on the middle layer at twist angle $\theta_{12}=\theta_{23}=1.46^\circ$. The system has the lowest energy when the supermoiré domain size is maximized by the strain at $\epsilon=1/\cos(\theta)-1=0.032\%$. \textbf{b}-\textbf{d}, the relaxed real-space lattice structure with middle layer biaxial strain $\epsilon$=0, 0.04\%, and 0.08\%, respectively. The red shading indicates the local moiré aperiodicity. The supermoiré domains show triangle shapes, with the AAA stacking sites (red circles) connected by domain walls. The orange and purple dots show the AA stacking regions of adjacent layer pairs. Inside the supermoiré domains, the relaxed lattice forms a moiré-periodic order (zoomed-in insets).
} 
    \label{fig:strain_relaxation}
\end{figure}

The results for twist angles $\theta_{12}=\theta_{23}=1.46^\circ$ are shown in Fig.~\ref{fig:strain_relaxation}. Figure~\ref{fig:strain_relaxation}\textbf{a} shows the total energy of the relaxed lattice configuration, including both stacking energy and elastic energy, as a function of the biaxial strain on the middle layer. The total energy is minimized at a strain where the supermoiré domain size is maximized ($\epsilon=1/\cos(\theta)-1=0.032\%$). This indicates that, in the absence of other extrinsic factors, the system prefers to relax to maximize the supermoiré domain size with a small cost of the intralayer elastic energy. Realistically, the global relaxation is affected by disorder and edge effects in the sample, which can explain the observation of enhanced but finite supermoiré domain size.

We plot the relaxed lattice structure at different middle layer strain strengths ($\epsilon=0, 0.04\%$, and 0.08\%) in Fig.~\ref{fig:strain_relaxation}\textbf{b}-\textbf{d}. Here, HTG relaxes into triangular supermoiré domains, with its AAA stacking sites denoted by red circles connected by domain walls. Different middle layer strain strength leads to very different supermoiré domain sizes. Inside each supermoiré domain, the orange and purple dots show the AA stacking regions of adjacent layer pairs, which form a honeycomb lattice, indicating that the system relaxes into moiré-periodic order. The moiré wavelength (determined by the periodicity of the orange/purple dots) is not affected by the strain. With this calculation, we directly confirm that the local moiré-periodic stacking order does not change under strain, whereas the size of the supermoiré domain drastically changes.

\subsection{g. Effect of strain on the extracted twist angle}

The relation $n_{s} = 4/A \approx 8\theta^{2}/\sqrt{3}a^{2}$ (see Methods) is precise when there is no strain or lattice relaxation. Realistically, strain and lattice relaxation lead to a small correction to this equation. For uniaxial strain of strength $\epsilon$ on one graphene layer, it expands the graphene lattice constant in one direction while contracts in the other direction. The resulting moiré pattern is therefore deformed to be anisotropic. Overall, the ratio of the change of the moiré unit cell size is on the order of $\epsilon$, which is typically below 0.7\% in graphene, and therefore is almost negligible when applying the above equation. Lattice relaxation effectively acts as a small amount of biaxial strain $\epsilon \approx (1/\cos\theta)-1$ on the middle layer graphene lattice. Its impact on the moiré unit cell area is also on the order of $\epsilon$. Thus, while strain and lattice relaxation can affect moiré unit cell area and the corresponding extracted angle, the effect is very small for our experimental parameters.

\section{4. Details of thermal cycling and sample changes between measurements}

\begin{figure}[t!]
    \renewcommand{\thefigure}{S\arabic{figure}}
    \centering
    \includegraphics[width=.7\columnwidth]{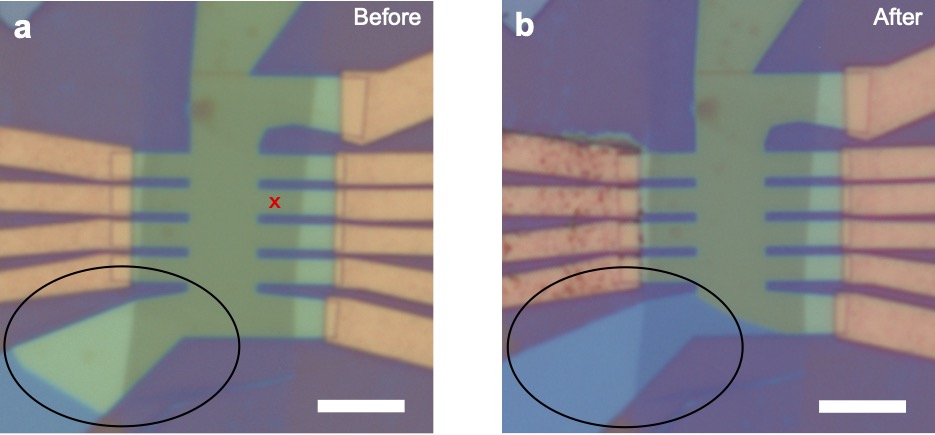}
    \caption{\textbf{Device Comparison between the two rounds of measurements}. \textbf{a}, \textbf{b}, Optical micrographs of the device before and after the sample change, respectively. The red ``x" in \textbf{a} denotes where the tip crashed and the black circles indicate where the top hBN layer was inadvertently removed. Scale bar: 5 $\mu$m.
} 
    \label{fig:device_before_after}
\end{figure}

In Fig.~2\textbf{d}-\textbf{f}, we show that the sample relaxed into larger and more symmetric supermoiré domains in a second round of measurements. This occurred after the sample was thermally cycled (i.e.~heated and cooled back down) and underwent other visible changes, detailed below.

The process was instigated by inadvertently crashing the SET tip on the sample. This occurred within the second contact leg from the top on the right side (red ``x" in Fig.~\ref{fig:device_before_after}\textbf{a}), away from the main region of measurement. The SET tip was shocked, necessitating a tip change that also required the sample to be removed from the cryostat. The sample was then brought to room temperature, optically investigated for any visible damage, and then reinserted into the cryostat for a second round of measurements after a new SET tip was fabricated. However, due to difficulty approaching the sample with the new tip, the sample experienced a total of three thermal cycles from cryogenic temperatures to room temperature. This included re-gluing it to the sample holder twice after shifting its position, with the glue cured at a temperature of 40-60 $^\circ$C overnight. During the optical investigation, we also noticed that a portion of the top hBN was removed from the device (circled region in Fig.~\ref{fig:device_before_after}). The exact cause of this is unknown. While we cannot definitively identify which of the above factor(s) contributed to the restructuring of the HTG supermoiré domains, these results suggest that intentional thermal annealing may be one possible method to enhance domain sizes. Lastly, we comment that the observation of more symmetric supermoiré domains indicates that the final strain configuration likely became more isotropic.

\section{5. Charging of the sample at a.c. timescales}

As mentioned in the main text, the a.c.~measurement modality of the SET may not allow carriers to fully equilibrate in highly resistive regions of the sample. This manifests when the resistance is large enough, e.g.~from a large energy gap, that $2\pi f \tau$ becomes non-negligible, where $f$ is the measurement frequency and $\tau$ is the $RC$ charging time constant~\cite{goodall1985capacitance,hunt2013massive,shi2020wse2}. In this case, the sample effectively acts more like an insulator in a.c.~measurements, and therefore appears to be more incompressible than in d.c.~measurements, resulting in ``a.c.~enhancement'' of incompressible signals. Here we provide additional evidence supporting the assertion that the degree of a.c.~enhancement can be traced back to dissipative components of the complex impedance, and thus encodes information about local resistance. 

To demonstrate that this is the correct conclusion, we characterize the sample charging under different conditions. We apply an a.c.~excitation $\delta V_{\rm{2D}}$ to the sample and vary its frequency $f_{\rm{2D}}$. The corresponding SET response $\delta I_{\rm{2D}}$ is directly proportional to the local electrostatic potential change induced by $\delta V_{\rm{2D}}$ and we record both its in-phase component $\delta I_{\rm{2D, x}}$ and its out-of-phase component $\delta I_{\rm{2D, y}}$. When the sample resistance is negligible, the impedance is purely capacitive and thus $\delta I_{\rm{2D}}$ is in phase with 
$\delta V_{\rm{2D}}$. However, in an incompressible state, the local resistance can become so large that it produces a non-zero out-of-phase signal. The ratio $\delta I_{\rm{2D, y}}/\delta I_{\rm{2D, x}}$, which is approximately $-2\pi f\tau$ from an effective circuit analysis, thus provides a qualitative probe for sample charging~\cite{goodall1985capacitance,hunt2013massive,shi2020wse2}. 

\begin{figure}[t!]
    \renewcommand{\thefigure}{S\arabic{figure}}
    \centering
    \includegraphics[width=.8\columnwidth]
    {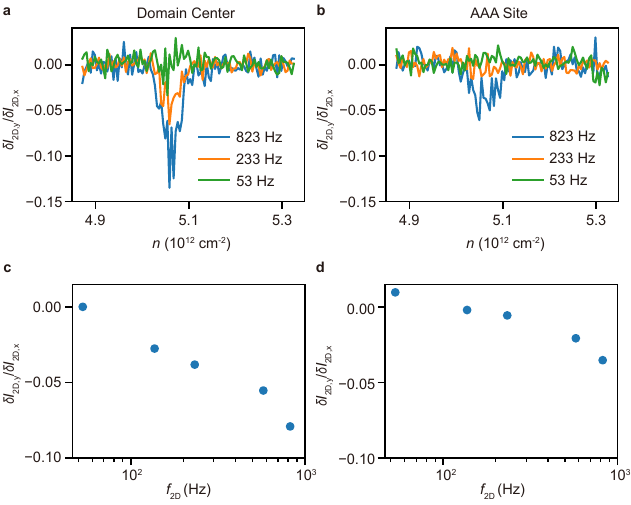}
    \caption{\textbf{Frequency dependence of $\delta I_{\rm{SET,2D}}$ near $\nu = 4$.} \textbf{a}-\textbf{b}, Ratio between the out-of-phase and the in-phase components of the SET current response $\delta I_{\rm{2D, y}}/ \delta I_{\rm{2D, x}}$ at the domain center (\textbf{a}) and the AAA site (\textbf{b}) as a function of $n$ near $\nu = 4$ for different a.c.~excitation frequencies $f_{\rm{2D}}$. \textbf{c}-\textbf{d}, $\delta I_{\rm{2D, y}}/ \delta I_{\rm{2D, x}}$ in the vicinity of $\nu = 4$ at the domain center (\textbf{c}) and the AAA site (\textbf{d}) as a function of $f_{\rm{2D}}$. For each frequency, $\delta I_{\rm{2D, y}}/\delta I_{\rm{2D, x}}$ is averaged over a density range of $4\times10^{10}~\mathrm{cm^{-2}}$ centered at $\nu = 4$.
} 
    \label{fig:frequency}
\end{figure}

We compare measurements near the incompressible $\nu = 4$ state at representative locations at a domain center (Fig.~\ref{fig:frequency}\textbf{a}) and at an AAA site (Fig.~\ref{fig:frequency}\textbf{b}) as a function of density and frequency. In all phase-sensitive measurements shown here, we adjust the phase so that $\delta I_{\rm{2D, y}} = 0$ in the compressible regime. In each location, $\delta I_{\rm{2D, y}}/\delta I_{\rm{2D, x}}$ decreases to negative values in the vicinity of $\nu=4$ (Fig.~\ref{fig:frequency}\textbf{a}-\textbf{b}), signaling insufficient charging at a.c.~time scales within the incompressible state. With decreasing a.c.~frequency, its magnitude $|\delta I_{\rm{2D, y}}/\delta I_{\rm{2D, x}}|$ drops, indicating better charging at lower frequencies. Moreover, comparing different locations at the same a.c. frequency, we find that $|\delta I_{\rm{2D, y}}/\delta I_{\rm{2D, x}}|$ is larger at the domain center than that at the AAA site (Fig.~\ref{fig:frequency}\textbf{a,b}), indicating a larger $RC$ time constant and worse charging at the domain center. In Fig.~\ref{fig:frequency}\textbf{c},\textbf{d}, we include measurements at additional intermediate $f_{\rm{2D}}$ and plot the value of $\delta I_{\rm{2D, y}}/\delta I_{\rm{2D, x}}$ in the immediate vicinity of the $\nu=4$ state at the domain center and the AAA site, respectively. The frequency dependence at each individual site (decreasing out-of-phase signal with decreasing frequency), the relative magnitude at each site (more prominent at the domain center), and the frequency at which the signal vanishes (lower at the domain center) all match well to theoretical expectations and further corroborate the conclusion that the degree of a.c.~enhancement is worse as the local resistance increases.

\section{6. Tip convolution simulations}

The experimental measurements reflect a convolution of the spatial modulation produced by the supermoiré domains in HTG and the spatial resolution of the SET tip. The latter is primarily limited by the finite tip size and its height above the sample. Typical SET tips have a metal island at their apex which is $\sim$100 nm in diameter and is held $\lesssim$100 nm above the sample during measurements. These length scales are of the same order as the size of the supermoiré features, and consequently, the effective spatial resolution of the SET must be taken into account to accurately model the expected experimental signatures. In this section, we provide a minimal model for the SET spatial resolution, accounting for the capacitive coupling between the tip and the sample. We demonstrate that, with realistic parameters, convolving the tip resolution kernel with the theoretically predicted spatial dependence of the chemical potential change $\Delta\mu_{4}$ across $\nu = 4$ matches the experimental results semi-quantitatively.

During measurement, voltages $V_{\rm{g}}$, $V_{\rm{2D}}$, and $V_{\rm{SET}}$ are individually applied to the graphite bottom gate, the sample, and the SET tip, respectively. Note that these voltages are electrochemical potentials, so the charge on the tip $q_{\rm{SET}}$ is given by
\begin{equation}
    q_{\rm{SET}} = C_{\rm{SET}}\phi_{\rm{SET}} + C_{\rm{ts}}\phi_{2\rm{D}} = C_{\rm{SET}}\left(V_{\rm{SET}} + \frac{\mu_{\rm{SET}}}{e}\right) + C_{\rm{ts}}\left(V_{2\rm{D}} + \frac{\mu}{e}\right),
\end{equation}
where $\phi_{\rm{SET}}$ and $\phi_{2\rm{D}}$ are the electrostatic potential of the SET tip and the sample, respectively, $C_{\rm{SET}}$ is the self-capacitance of the SET tip, $C_{\rm{ts}}$ is the tip-sample capacitance, $\mu_{\rm{SET}}$ is the chemical potential of the tip, $\mu$ is the chemical potential of the sample, and $e$ is the (positive) elementary charge.

If the sample has a spatially inhomogeneous chemical potential, for example from either local doping or domain formation, obtaining the charge induced on the tip requires spatial integration:
\begin{equation}
    q_{\rm{SET}} = C_{\rm{SET}}\left(V_{\rm{SET}} + \frac{\mu_{\rm{SET}}}{e}\right) + \iint_{s}C_{\rm{ts}}(x,y){\rm{d}}x{\rm{d}}y\left[V_{2\rm{D}} + \frac{\mu(x,y)}{e}\right],
\end{equation}
where the spatial integral is over the $xy$ plane on the sample and we have assumed the tip is at the origin. Because we employ a feedback mechanism on $V_{2\rm{D}}$ to maintain a constant SET working point, the charge on the SET island remains constant as we change the carrier density (and thus chemical potential) in the sample: 
\begin{equation}
    0 = \delta q_{\rm{SET}} = \iint_{s}C_{\rm{ts}}(x,y){\rm{d}}S\left[\Delta V_{2\rm{D}} + \frac{\Delta\mu(x,y)}{e}\right],
\end{equation}
Therefore, in the presence of chemical potential inhomogeneity across the sample, the d.c.~measurement modality, which yields $\Delta V_{\rm{2D}}$ while maintaining $\delta q_{\rm{SET}} = 0$, results in a spatial convolution of the local chemical potential change $\Delta \mu(x,y)$, with the convolution kernel being $C_{\rm{ts}}(x,y)$:
\begin{equation}
   e\Delta V_{\rm{2D}} = -\frac{\iint_{s}C_{\rm{ts}}(x,y)\Delta\mu(x,y){\rm{d}}S}{\iint_{s}C_{\rm{ts}}(x,y){\rm{d}}S},
\end{equation}

\begin{figure}[t!]
    \renewcommand{\thefigure}{S\arabic{figure}}
    \centering
    \includegraphics[width=0.85\columnwidth]{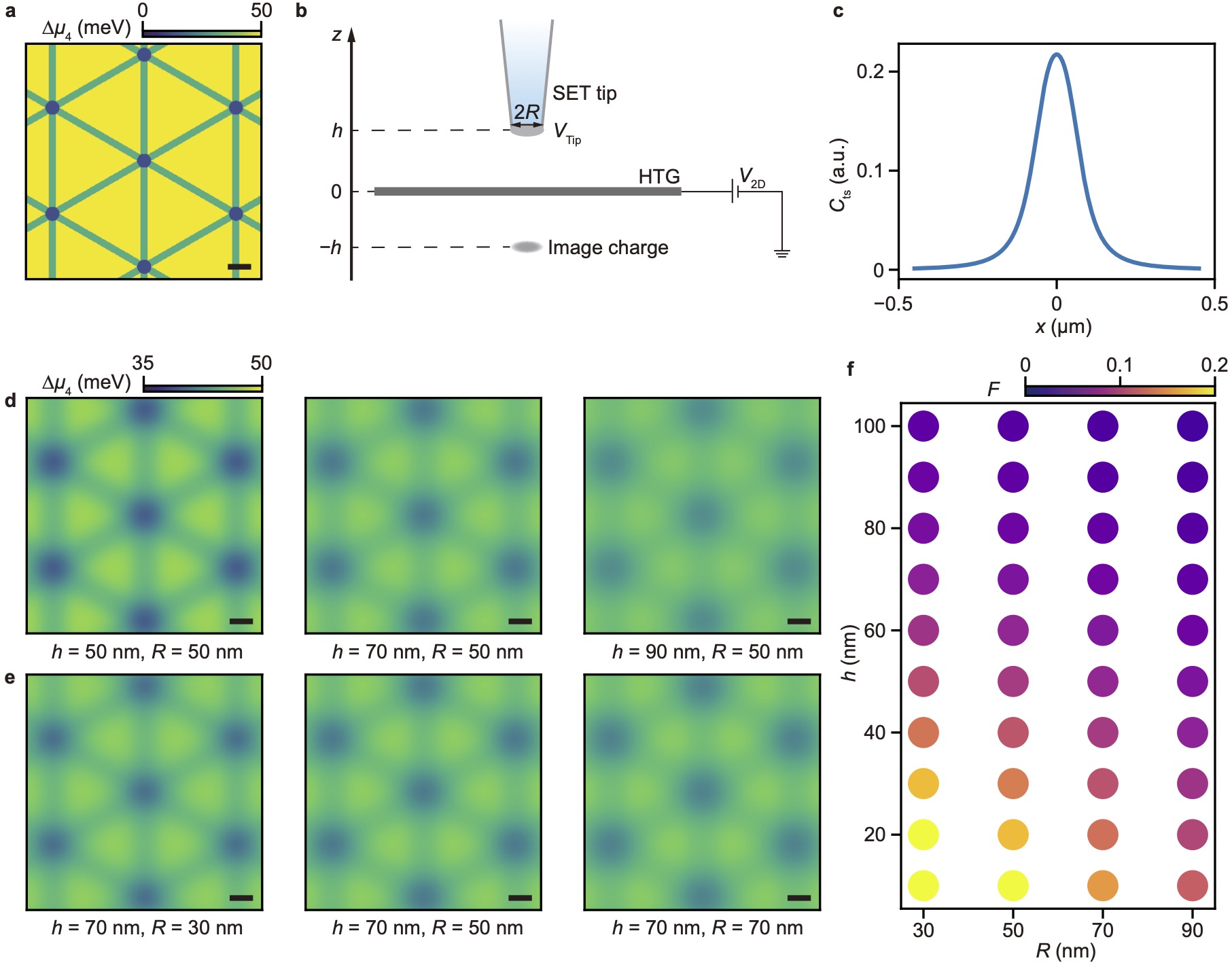}
    \caption{\textbf{Relaxed supermoiré structure and tip-convolution simulation} \textbf{a}, Map of chemical potential change $\Delta\mu_{\rm{4}}$ of the modeled reconstructed supermoiré structure. Relevant parameters are described in the supplementary text. Scale bar: 100~nm. \textbf{b}, Schematic of the electrostatic model, whose parameters are the tip radius $R$ and its height $h$ above the HTG. \textbf{c}, Line cut of normalized $C_{\rm{ts}}(x,y)$ through $y=0$, with $h = 70~\rm{nm}$ and $R = 50~\rm{nm}$. \textbf{d}-\textbf{e}, Spatial map of the tip-convolved $\Delta\mu_{\rm{4}}$ based on the supermoiré structure in \textbf{a} as a function of $h$ (\textbf{d}) and $R$ (\textbf{e}). All panels share the same color bar. Scale bar: 100~nm. \textbf{f}, Visibility $F$ of the convolved supermoiré modulation as a function of $R$ and $h$.}
    \label{fig:sim_eg}
\end{figure}

To accurately simulate the experimental supermoiré features expected from the d.c.~measurement modality, we need to model (i) the tip convolution kernel $C_{\rm{ts}}(x,y)$ which is mainly determined by electrostatics, and (ii) the spatial modulation of chemical potential $\Delta \mu(x,y)$ produced by the HTG supermoiré domains.

For (i), the coefficient $C_{\rm{ts}}(x,y)$ describes the capacitance between a tip held at $(x,y,z)=(0,0,h)$ and a unit area at $(x,y,0)$ on the sample, which is purely geometric and does not depend on sample details except the sample being conductive (valid above and below $\nu=4$, which we use to determine $\Delta\mu_4$). Therefore, without loss of generality, $C_{\rm{ts}}(x,y)$ can be determined by considering the following electrostatic model (Fig.~\ref{fig:sim_eg}\textbf{b}): the SET tip is modeled as a metal disk with radius $R$, centered at $(0,0,h)$, while the sample, when compressible, can be approximated as a homogeneous infinite metal plane at $z = 0$. When the disk is held at electrostatic potential $\phi = V$, and the sample plane grounded at $\phi = 0$, the surface charge density $\sigma(x,y)$ induced on the sample plane is
\begin{equation}
   \sigma(x,y) = C_{\rm{ts}}(x,y)V.
\end{equation}
Therefore, solving the electric field on the sample plane $E(x,y,0)$ directly gives $C_{\rm{ts}}(x,y)$: 
\begin{equation}
   C_{\rm{ts}}(x,y) = \frac{\sigma(x,y)}{V} = \frac{\epsilon_0}{V} E(x,y,0) \propto E(x,y,0).
\end{equation}

We proceed to solve the aforementioned electrostatic model. From the principle of image charges, the electric field above the sample plane is equivalent to that of the tip disk plus an image charge, which is an oppositely charged, equal sized disk located at $(0, 0, -h)$. The electrostatic potential can be solved as \cite{10.1093/qjmam/2.4.428, 10.1119/1.17668, 10.1098/rspa.2016.0574}
\begin{equation}\label{phi}
   \phi(x,y,z) \propto \frac{2R}{\pi}\cdot\mathrm{Re}\int_0^1 f(t)dt\left[ \frac{1}{\sqrt{x^2 + y^2 + (|z+h|-iRt)^2}} - \frac{1}{\sqrt{x^2 + y^2 + (|z-h|-iRt)^2}}\right],
\end{equation}
and consequently the electric field on the sample plane is
\begin{equation}\label{E-field}
   E(x,y,0) = -\frac{\partial\phi}{\partial z}\Big|_{z=0} \propto \frac{4R}{\pi}\cdot\mathrm{Re}\left[\int_0^1 dt \frac{(h-iRt)f(t)}{(x^2 + y^2 + (h-iRt)^2)^{\frac{3}{2}}}\right].
\end{equation}

The function $f(t)$ is the solution to Love's equation~\cite{10.1093/qjmam/2.4.428}, and we calculate its asymptotic expansion under the limit $\eta=2h/R \to 0$ up to $\eta^3$ terms according to~\cite{PhysRevResearch.2.013289}. This level of approximation leads to a total capacitance that is $\lesssim 1\%$ different from that obtained upon including the next term in the expansion in the parameter range we have used.
Substituting $f(t)$ into Eq.~(\ref{E-field}), we can numerically calculate $C_{\rm{ts}}(x,y)$ as a function of tip radius $R$ and height $h$ up to a proportionality factor and normalized to the integral $\iint C_{\rm{ts}}(x,y) dxdy$, which is sufficient for our purpose of convolution. A line cut at $y = 0$ of $C_{\rm{ts}}(x,y)$ with $h = 70$ nm and $R = 50$ nm is shown in Fig.~\ref{fig:sim_eg}\textbf{c}, which provides a representation of the effective spatial resolution of the measurement.

For task (ii), we rely on a simplified picture of the relaxed supermoiré structure to construct the $\Delta \mu(x,y)$ map before tip convolution. Three components are taken into account: domains, domain walls, and AAA sites. Using the $\nu = +4$ gap as an example, we extract the chemical potential change $\Delta\mu$ across $\nu = +4$ (density range $n = 4.785\text{--}4.985\times10^{12}\ \rm{cm^{-2}}$) for each component from the theoretically calculated $\mu(n)$ (Fig.~\ref{fig:finite_D_band}\textbf{d}-\textbf{f}).
We estimate the domain wall thickness to be $t_{\rm{dw}}=30$~nm and the AAA site radius $R_{\rm{AAA}}=30$~nm~\cite{devakul2023magic}. The supermoiré wavelength, i.e.~the distance between AAA sites, is set to $451\ \mathrm{nm}$, which approximately matches the experimentally observed supermoiré size. The corresponding map of $\Delta\mu(x, y)$ of the relaxed supermoiré structure, is shown in Fig.~\ref{fig:sim_eg}\textbf{a} and main text Fig.~3\textbf{h}.

To explore how the tip radius and height affect the visibility of spatial modulations in experiment, we convolve the $\Delta\mu_{4}$ of supermoiré structure in Fig.~\ref{fig:sim_eg}\textbf{a} with $C_{\rm{ts}}(x,y)$ for varying $R$ and $h$ (Fig.~\ref{fig:sim_eg}\textbf{d},\textbf{e}). Qualitatively, all maps exhibit similar variations, but the contrast of the spatial modulations rapidly drops with increasing $h$ and $R$. To quantify this change, we define the visibility $F$ of the supermoiré structure as 
\begin{equation}
    F = \frac{\Delta\mu_{\rm{max}} - \Delta\mu_{\rm{min}}}{\Delta\mu_{\rm{max}} + \Delta\mu_{\rm{min}}}.
\end{equation}
In Fig.~\ref{fig:sim_eg}\textbf{f}, we show how $F$ varies as a function of $h$ and $R$ over a more comprehensive parameter range. For the tip-convolution simulation shown in Fig.~3\textbf{i} of the main text, we set $h$ = 70 nm and $R$ = 50 nm, which are realistic parameters and match with the experimentally observed visibility $F\sim0.059$ for the $\Delta\mu_{\rm{4, d.c.}}$ map in Fig.~3\textbf{b}.

\bibliographystyle{apsrev4-1}
\bibliography{references.bib}